\documentclass[12pt]{article} 

\usepackage{amsmath,amssymb,amsfonts}
\usepackage{psfrag}
\usepackage{color}
\definecolor{darkblue}{rgb}{0.1,0.1,.7}
\definecolor{purple}{rgb}{0.6,0,0.6}
\definecolor{orange}{rgb}{0.9,0.6,0}
\usepackage[colorlinks, linkcolor=darkblue, citecolor=darkblue, urlcolor=darkblue, linktocpage]{hyperref} 
\usepackage[square, comma, compress,numbers]{natbib}
\usepackage[]{amsmath}
\usepackage[]{graphicx}
\usepackage[]{latexsym}
\usepackage{geometry}
\usepackage{amscd}
\usepackage[all,cmtip]{xy}
\usepackage{mathrsfs}
\usepackage[margin=10pt,font=small,labelfont=bf]{caption}
\usepackage{dsdshorthand}
\usepackage{changepage}
\usepackage{tikz}
\usetikzlibrary{shapes,arrows}
\usetikzlibrary{arrows,automata}

\tikzset{
    state/.style={
           rectangle,
           rounded corners,
           draw=black, very thick,
           minimum height=2em,
           inner sep=2pt,
           text centered,
           },
}

\usepackage{subcaption}

\def\zb{\overline z}
\def\hb{\overline h}

\newcommand\Disc{\textrm{Disc}}
\newcommand\PFQ[5]{{}_#1 F_#2\left(\genfrac{}{}{0pt}{}{#3}{#4};#5\right)}

\numberwithin{equation}{section}

\begin{document}
\pagenumbering{gobble}

\vspace*{-.6in} \thispagestyle{empty}
\begin{flushright}
\hfill{\tt CALT-TH-2018-049 }\\
\hfill{\tt PUPT-2573}
\end{flushright}
\vspace{.2in} {\Large
\begin{center}
{\bf Bootstrapping the 3d Ising model \\ at finite temperature \\\vspace{.1in}}
\end{center}
}
\vspace{.2in}
\begin{center}
{\bf 
Luca Iliesiu$^{1}$,
Murat Kolo\u{g}lu$^{2}$,
David Simmons-Duffin$^{2}$} 
\\
\vspace{.2in} 
$^1${\it Department of Physics, Princeton University, Princeton, NJ 08540, USA\/} \\
$^2${\it Walter Burke Institute for Theoretical Physics, Caltech, Pasadena, California 91125\/}
\end{center}

\vspace{.2in}

\begin{abstract}
We estimate thermal one-point functions in the 3d Ising CFT using the operator product expansion (OPE) and the Kubo-ÐMartin-ÐSchwinger (KMS) condition. Several operator dimensions and OPE coefficients of the theory are known from the numerical bootstrap for flat-space four-point functions. Taking this data as input, we use a thermal Lorentzian inversion formula to compute thermal one-point coefficients of the first few Regge trajectories in terms of a small number of unknown parameters. We approximately determine the unknown parameters by imposing the KMS condition on the two-point functions $\<\sigma\sigma\>$ and $\<\epsilon\epsilon\>$. As a result, we estimate the one-point functions of the lowest-dimension $\mathbb Z_2$-even scalar $\epsilon$ and the stress energy tensor $T_{\mu \nu}$. Our result for $\<\sigma\sigma\>$ at finite-temperature agrees with Monte-Carlo simulations within a few percent, inside the radius of convergence of the OPE.
\end{abstract}

\newpage

\tableofcontents

\pagenumbering{arabic}

\section{Introduction}
\label{sec:intro}

In \cite{Iliesiu:2018fao}, we initiated a study of conformal field theories at finite (i.e.\ nonzero) temperature in $d>2$ dimensions, using techniques from the conformal bootstrap. At finite temperature, the operator product expansion (OPE) can still be used to reduce $n$-point correlators to sums of $n{-}1$-point correlators. However, an important new ingredient at temperature $T=1/\b$ is that non-unit operators can have nonzero one-point functions $\<\cO\>_\beta$. For example, the thermal one-point function of the stress tensor $\<T_{\mu\nu}\>_\b$ encodes the free-energy density.

Thermal one-point functions are constrained by a type of ``crossing-equation" first written down by El-Showk and Papadodimas \cite{ElShowk:2011ag}. They noted that the Kubo-ÐMartin-ÐSchwinger (KMS) condition for thermal two-point functions is not manifestly consistent with the OPE, and this leads to constraints on CFT data. An efficient way to study these constraints is to use the thermal Lorentzian inversion formula developed in \cite{Iliesiu:2018fao}, which is an analog of Caron-Huot's Lorentzian inversion formula for zero-temperature four-point functions \cite{Caron-Huot:2017vep, Simmons-Duffin:2017nub, Kravchuk:2018htv}.

In this work, we apply these ideas to estimate thermal one- and two-point functions in a strongly-coupled conformal field theory in $d=3$ dimensions: the 3d Ising CFT. Physically, this theory describes the 2+1-dimensional quantum transverse field Ising model at nonzero temperature, and the 3-dimensional statistical Ising model with a periodic direction of length $\b$ (both at criticality).\footnote{Note that the temperature we discuss in this work is not related to the ``temperature" of the statistical Ising model that determines the spin-spin coupling. The latter quantity is set to its critical value.} Besides its physical interest, an advantage of studying the 3d Ising CFT is that we can leverage a wealth of information about its zero-temperature OPE data from the conformal bootstrap \cite{ElShowk:2012ht,El-Showk:2014dwa,Kos:2014bka,Kos:2016ysd,Simmons-Duffin:2016wlq}.  The 3d Ising CFT is a case where Monte-Carlo (MC) techniques are also very efficient for computing some finite-temperature observables \cite{Hasenbusch:2011yya}. However, we believe it is worthwhile to develop bootstrap-based approaches. One might hope to eventually apply these approaches to theories that are more difficult to study with MC, like fermionic theories, or non-Lagrangian CFTs.

The thermal crossing equation of El-Showk and Papadodimas is difficult to study for two reasons. Firstly, it does not enjoy the positivity conditions that are important for rigorous numerical bootstrap techniques to work \cite{Rattazzi:2008pe,Hellerman:2009bu,Poland:2011ey,Paulos:2014vya,Simmons-Duffin:2015qma,Poland:2018epd}. Thus, we will not be able to compute rigorous bounds on thermal data and will have to content ourselves with estimates. Our rough strategy is to truncate the thermal crossing equation and approximate it by a finite set of linear equations for a finite set of variables. In spirit, this is similar to the ``severe truncation" method initiated by Gliozzi \cite{Gliozzi:2013ysa, Gliozzi:2014jsa} and applied with some success in the boundary/defect bootstrap \cite{Liendo:2012hy,Gliozzi:2015qsa,Rastelli:2017ecj, Billo:2016cpy,Gadde:2016fbj,Liendo:2016ymz,Lauria:2017wav,Lemos:2017vnx}.

However, a second difficulty is that the thermal crossing equation converges more slowly than the crossing equation for flat-space four-point functions. Thus, na\"ive ``severe truncation" is doomed to fail, and we need a more sophisticated approach. We will use the thermal Lorentzian inversion formula and large-spin perturbation theory to estimate the behavior of a few families of operators (specifically, the first few Regge trajectories) in terms of a small number of unknown parameters. This reduces the number of unknowns in the crossing equations and allows them to be solved approximately by a least-squares fit. 

In section~\ref{sec:review}, we review the conformal bootstrap at finite temperature, following \cite{Iliesiu:2018fao}, together with some features of the spectrum of the 3d Ising CFT \cite{Simmons-Duffin:2016wlq} that play an important role in our calculation. In section~\ref{sec:method-and-results}, we outline our overall strategy and summarize the results. As a check, we perform an MC simulation of the 3d critical Ising model and find agreement with our determination of $\<\s\s\>_\b$ to within statistical error, inside the regime of convergence of the OPE. Section~\ref{sec:details} presents the details of our bootstrap-based calculation.  The most complicated step is the estimation of thermal one-point coefficients for subleading Regge trajectories, which we perform by adapting the ``twist-Hamiltonian" procedure of \cite{Simmons-Duffin:2016wlq}.

\section{Review}
\label{sec:review}

\subsection{The thermal bootstrap}
\label{sec:thermal-review}
 
A CFT at nonzero temperature $T$ can equivalently be thought of as living on the space $S^1_\beta \times \R^{d-1}$, where $\b=1/T$ is the length of the thermal circle. This space is conformally flat, so one can compute finite-temperature correlators using the OPE, just as in flat space. However, an important difference compared to flat-space is that the thermal circle introduces a scale, and as a result operators can have nonzero one-point functions. Symmetries imply that the only operators with nonzero one-point functions are primary even-spin traceless symmetric tensors $\cO^{\mu_1\cdots\mu_J}$. For such operators, we have
\be
\label{eq:thermalonept}
\<\cO^{\mu_1\cdots \mu_J}(x)\>_{S_\b^1 \x \R^{d-1}} &= 
\frac{b_\cO}{\b^{\De}} (e^{\mu_1} \cdots e^{\mu_J} - \textrm{traces}),
\ee
where $\De$ is the dimension of $\cO$, $e^\mu$ is a unit vector in the $S^1$ direction, and $b_\cO$ is a dynamical constant.

Consider a two-point function of a real scalar primary $\f$ at finite temperature:
\begin{align} 
g(\tau, \bx) = \<\f(\tau, \bx) \f(0)\>_{S_\b^1 \x \R^{d-1}}.
\end{align}
Here, we introduced coordinates $x=(\tau,\bx)$, where $\tau\in[0,\b)$ and $\bx\in \R^{d-1}$.  Assuming $|x|=(\tau^2 + \bx^2)^{1/2}<\b$, this two-point function can be evaluated using the OPE:
\be 
\label{eq:two-pt-function-expansion}
g(\tau,\bx) &= \sum_{\cO\in \f\x \f} \frac{a_\cO^{\langle \phi \phi\rangle}}{\b^\De} C_J^{(\nu)}\left(\frac{x\.e}{|x|}\right) |x|^{\De-2\De_\f}\,,\nn\\
 a_\cO^{\langle \phi \phi\rangle} &\equiv f_{\f\f\cO} b_\cO \frac{J!}{2^J (\nu)_J}.
\ee
Here, $\cO$ runs over primary operators appearing in the $\phi\times \phi$ OPE, with OPE coefficients $f_{\f\f\cO}$. $\Delta$ is the scaling dimension of $\cO$, $J$ is its spin, and $\nu = (d-2)/2$. We call each term in (\ref{eq:two-pt-function-expansion}) a ``thermal block." The thermal one-point coefficient $b_\cO$ is defined in (\ref{eq:thermalonept}), and we have defined the thermal coefficients $a_\cO^{\<\f\f\>}$ for later convenience.

For simplicity, we set $\b=1$ in what follows.
Let us use $d-1$-dimensional rotational invariance to set $\bx=(x,0,\dots,0)\in \R^{d-1}$ and introduce the coordinates
\be
z = \tau + i x,\quad \bar z = \tau - i x.
\ee
Note that $z,\bar z$ are complex conjugates in Euclidean signature.

The two-point function $g(\tau,\bx)$ is invariant under $\tau\to 1-\tau$. In the language of thermal physics, this is the KMS condition, and it is furthermore obvious from the geometry of $S^1_\b\x\R^{d-1}$. However, the OPE expansion (\ref{eq:two-pt-function-expansion}) is not manifestly invariant under $\tau\to 1-\tau$. This leads to a nontrivial crossing equation that constrains  thermal one-point functions $b_\cO$ in terms of scaling dimensions and OPE coefficients \cite{ElShowk:2011ag}. In terms of $z$ and $\bar z$, the crossing equation/KMS condition is
\be
\label{eq:KMS-condition}
g(z,\bar z) &= g(1-z,1-\bar z).
\ee
Here, we have also used that $g$ is invariant under $x\to -x$.

The coefficients $a_\cO^{\<\f\f\>}$ can be encoded in a function $a^{\<\f\f\>}(\De,J)$ that is meromorphic for $\De$ in the right-half-plane, with residues of the form
\be
a^{\<\f\f\>}(\De,J) &\sim - \frac{a_\cO^{\<\f\f\>}}{\De-\De_\cO}.
\ee
In \cite{Iliesiu:2018fao}, we showed that such a function can be obtained from a ``thermal Lorentzian inversion formula"
\be
a^{\langle \phi \phi\rangle}(\De,J) 
&=
(1+(-1)^J) K_J \int_0^1 \frac{d\bar z}{\bar z} \int_1^{1/\bar z} \frac{dz}{z} (z \bar z)^{\De_\f-\frac \De 2-\nu}(z-\bar z)^{2\nu} F_J\p{\sqrt{\frac {\bar z} {z}}} \Disc[g(z,\bar z)]\nn\\ &\quad+\theta(J_0-J) a_{\rm arcs}^{\langle \phi \phi\rangle}(\De,J)\,.
\label{eq:inversionformulahigherd}
\ee
Here $z,\bar z$ are treated as independent real variables, which means that the integral is over a Lorentzian regime $x\to -i x_L$.
The first term contains the discontinuity
 \be
\textrm{Disc}[g(z,\bar z)] &\equiv \frac{1}{i} \p{g(z+i\e,\bar z) - g(z-i\e,\bar z)}\, ,
\ee
and the functions $K_J $ and $F_J(w)$ are given by
\be
	K_J &\equiv 
 \frac{\G(J+1)\G(\nu)}{4\pi \G(J+\nu)}\, ,
\\	F_J(w) &= w^{J+d-2}{}_2F_1\p{J+d-2,\frac d 2 - 1, J+\frac d 2, w^2}\, .
\ee
The second line in (\ref{eq:inversionformulahigherd}) represents additional contributions that are present when $J<J_0$, where $J_0$ controls the behavior of the two-point function in a Regge-like regime. We argued in \cite{Iliesiu:2018fao} that $J_0<0$ for the 3d Ising CFT. In this work, we assume this is true and ignore these contributions.

\subsubsection{Large-spin perturbation theory}

The thermal inversion formula \eqref{eq:inversionformulahigherd} becomes particularly powerful in conjunction with the KMS condition (\ref{eq:KMS-condition}).

Let us call \eqref{eq:two-pt-function-expansion} the $s$-channel OPE, which in our new coordinates is an expansion around $z=\zb =0$ and has the region of convergence
\be 
\text{$s$-channel OPE: } \qquad |z|,|\bar z|<1\,.
\ee
 By the KMS condition, the two-point function admits another expansion around $z=\bar z = 1$, which we call the $t$-channel:
\begin{align}
g(z,\bar z) &= \sum_{\mathcal{O} \in \phi \times \phi} a_{\mathcal{O}}^{\langle \phi \phi\rangle} ((1-z)(1-\bar z))^{\frac{\Delta_{\mathcal{O}}}{2} -\Delta_\phi} C_{\ell_{\mathcal{O}}}^{(\nu)} \left(\frac{1}{2} \left(\sqrt{\frac{1-z}{1-\bar z}}+ \sqrt{\frac{1-\bar z}{1-z}} \right) \right) \label{eq:2pt function t-channel} \,.
\end{align}
Its region of convergence is given by:
\be 
\text{$t$-channel OPE: } \qquad |1-z|,|1-\bar z|<1\,.
\ee
We can insert the $t$-channel OPE into the inversion formula \eqref{eq:inversionformulahigherd} to find expressions for thermal coefficients in the $s$-channel. 
In this way, we uncover non-trivial relations between the thermal coefficients of different operators in the theory.

The integral in the inversion formula \eqref{eq:inversionformulahigherd} is within the region of convergence of the $t$-channel OPE for $1 \leq z < 2$, but for $z\geq2$ it exits this region. 
Corrections to the residues of $a(\Delta, J)$ coming from the region $z\geq2$ are exponentially suppressed in $J$. Thus, the $t$-channel OPE encodes the all-orders expansion in powers of $1/J$ for thermal one-point coefficients.

Let us review how poles and residues of $a(\Delta, J)$ arise from the thermal inversion formula. As an example, we study the poles and residues contributed by a single $t$-channel block. Individual $t$-channel blocks contribute poles at double-twist locations $\De=2\De_\f+2n+J$ \cite{Iliesiu:2018fao}. A similar phenomenon occurs in the flat-space lightcone bootstrap, where individual $t$-channel blocks again contribute to OPE data of double-twist operators. To obtain poles at other locations, one must sum infinite families of $t$-channel blocks before plugging them into the inversion formula. (We will see several examples below.) Nevertheless, individual $t$-channel blocks provide an important example that will be a building block for later calculations.

Poles in $\De$ come from the region $\bar z \sim 0$. Therefore, when computing residues one can simply replace the upper bound of the $z$ integral with $1/\bar z \sim \infty$. However, the range of the $z$ integral must then be artificially restricted to $z_{\max} =2$ when plugging in the $t$-channel expansion, in order for the $z$ integral to fully be within the region of OPE convergence. This restriction is essentially an approximation that discards corrections that die exponentially in $J$.

The residues are determined by a one-dimensional integral over $z$. To see this, we first expand the function $F_J\p{\sqrt{ {\bar z}/{z}}}$ in $\zb$ in the inversion formula,
\begin{align}
a^{\langle \phi \phi\rangle}(\Delta,J) = (1+(-1)^J) K_J \int_0^1 \frac{d\bar z}{\bar z} \int_1^{1/\bar z} \frac{dz}{z} \sum_{r=0}^\infty q_r(J) z^{\Delta_\phi-\bar h-r}\bar z^{\Delta_\phi-h+r} {\rm Disc}[g(z,\bar z)]\,, \label{eq:inversion series expansion}
\end{align} 
where the coefficients $q_r(J)$ are
\begin{align}
q_r(J) &\equiv  
(-1)^r \frac{(J+2 r)}{J} \frac{ (J)_{r} (-r+\nu +1)_r}{r! (J+\nu +1)_r}\,,\label{eq:def inversion series coefficients}
\end{align}
and we have rewritten the inversion formula in terms of the quantum numbers
\be
\label{eq:h-bar-h-def}
h = \frac{\Delta-J}2\,, \qquad \bar h = \frac{\Delta+J}2\,.
\ee

The $t$-channel OPE can also be expanded in a power series in $(1-z)$ and $(1-\overline{z})$, 
\be 
g(z, \bar z) = \sum_{\mathcal{O} \in \phi \times \phi} a_{\mathcal{O}}^{\langle \phi \phi\rangle} \sum_{s=0}^{\ell_{\mathcal{O}}} p_s(\ell_\cO) (1-z)^{h_{\mathcal{O}}-\Delta_\phi +s} (1-\bar z)^{\bar h_{\mathcal{O}} -\Delta_\phi -s},
\ee
where
\begin{align}
p_s(\ell) &\equiv \frac{\Gamma(\ell -s +\nu) \Gamma(s+\nu)}{\Gamma(\ell-s+1)\Gamma(s+1)}\frac{1}{ \Gamma(\nu)^2}
%\\ &
=\frac{1}{4\pi K_{\ell}} \frac{(\ell+\nu)_{-s}}{(\ell+1)_{-s}} \binom{\nu+s-1}{s}\,. \label{eq:def t-channel series coefficients}
\end{align}
The $h_\cO$ and $\bar h_\cO$ are the quantum numbers defined by (\ref{eq:h-bar-h-def}) for each $\cO$ appearing in the OPE. Plugging in the term corresponding to an individual $\cO$ from the $t$-channel OPE into the inversion formula \eqref{eq:inversionformulahigherd}, we find\footnote{We assume that $J$ is larger than $J_0$, so that the arcs do not contribute. As mentioned above, we expect $J_0<0$ in the 3d Ising CFT, so the arcs don't contribute to the pole of any local operator.}
\begin{align}
a^{{\langle \phi \phi\rangle},\,(\cal O)}(\Delta,J) &\approx (1+(-1)^J) K_J \int_0^1 \frac{d\bar z}{\bar z} \int_1^{z_{\text{max}}} \frac{dz}{z} \sum_{r=0}^\infty q_r(J) z^{\Delta_\phi-\bar h-r}\bar z^{\Delta_\phi-h+r} \nonumber
\\ & \qquad \times {\rm Disc}\left[a_{\cal O}^{\langle \phi \phi\rangle} \sum_{s=0}^{\ell_{\cal O}} p_s(\ell_\cO) (1-z)^{h_{\mathcal{O}}-\Delta_\phi +s} (1-\bar z)^{\bar h_{\mathcal{O}} -\Delta_\phi -s}
\right]
\nn\\ &=  a_{\cal O}^{{\langle \phi \phi\rangle}} (1+(-1)^J) K_J   \sum_{r=0}^\infty \sum_{s=0}^{\ell_{\cal O}} q_r(J) p_s(\ell_\cO) \frac{\Gamma(1+\bar h_\cO - \Delta_\phi-s) \Gamma(\Delta_\phi+r-h) }{\Gamma(\bar h_\cO -h +1-s+r)} \nonumber
\\ & \qquad \times   2\pi S_{h_\cO-\Delta_\phi+s,\Delta_\phi-r}(\bar h)\,,\label{eq:single t-chan op inversion}
\end{align}
Here, the superscripts $a^{{\langle \phi \phi\rangle},\,(\cal O)}$ indicate that we are studying thermal coefficients for $\<\f\f\>$, focusing on the contribution of the $t$-channel operator $\cO$.
To go from the first equation to the second equation above we have performed the $\zb$ integral and defined the function $S_{h_\cO-\Delta_\phi+s,\Delta_\phi-r}(\bar h)$ as
\begin{align}
S_{c,\Delta}(\bar h) &= \frac{\sin(-\pi c)}{\pi} \int_1^{z_\text{max}} \frac{dz}{z} z^{\Delta-\bar h} (z-1)^c 
\nn\\ &= \frac{1}{\Gamma(-c)} \frac{\Gamma(\bar h-\Delta-c)}{\Gamma(\bar h-\Delta+1)} - \frac{1}{\Gamma(-c)\Gamma(1+c)} B_{1/z_\text{max}}(\bar h-\Delta-c,1+c)\,. \label{eq:def S}
\end{align}
Here $B_{1/z_\text{max}}(\bar h-\Delta-c,1+c)$ is the incomplete beta function, which decays as $z_\text{max}^{-\bar h} \sim z_\text{max}^{-J}$ at large $\bar h$.

 Note that in \eqref{eq:single t-chan op inversion} the $\bar z$-integral has generated poles at double-twist locations $\Delta = 2\Delta_\phi + 2n + J$, coming from the factors $\G(\De_\f+r-h)$. 
Taking the residue of \eqref{eq:single t-chan op inversion}, we get the contribution of the operator $\cO$ to the $[\f\f]_n$ families 
\begin{align}
& a_{[\f\f]_{n}}^{{\langle \phi \phi\rangle},\,(\cO)} ( J) = -\Res\limits_{\Delta = 2\Delta_\phi +2n+J} a^{{\langle \phi \phi\rangle},\,(\cO)} (\Delta,J)
\nn\\ &\quad = a_\cO^{\langle \phi \phi\rangle} (1+(-1)^J) 4\pi K_J \frac{d\bar h}{dJ} \sum_{r=0}^{n} \sum_{s=0}^{\ell_{\cal O}} q_r(J) p_s(\ell_\cO) (-1)^{n-r} \binom{\bar h_\cO-\Delta_\phi -s}{n-r} 
S_{h_\cO-\Delta_\phi+s,\Delta_\phi-r}(\bar h). \label{eq:residue of single t-chan op inversion}
\end{align}
For double-twist operators $[\f\f]_n$, we have $\bar h = \De_\f+n + J$. The Jacobian factor $\frac{d\bar h}{dJ}$ takes into account the leading correction to (\ref{eq:residue of single t-chan op inversion}) when we additionally allow $[\f\f]_n$ to have anomalous dimensions.

The function $S_{h_\cO-\De_\f+s,\De_\f-r}(\bar h)$ can be expanded in large $\bar h$ (equivalently large $J$) as
 \begin{align}
S_{h_\cO - \Delta_\phi+s,\Delta_\f - r}(\bar h) = \frac{1}{\Gamma(-h_\cO + \Delta_\phi - s)} \frac{1}{\bar h^{h_\cO - \Delta_\phi+s+1}} + \cO\left(\frac{1}{\bar h^{h_\cO - \Delta_\phi+s+2}}\right). \label{eq:S leading asymptotics}
\end{align} 
Thus, we see that the contribution of the $t$-channel operator $\cO$ dies at large $J$ at a rate controlled by the half-twist $h_\cO=\tau_\cO/2$. The unit operator has the lowest twist in any unitary theory, and thus gives the leading contribution at large $J$. A second important contribution comes from the stress tensor $\cO=T_{\mu\nu}$, which gives a universal contribution proportional to the free energy density. In general, by including successively higher-twist contributions in the $t$-channel, we can build up a perturbative expansion for thermal coefficients in $1/J$. We will review this large-spin perturbation theory of the thermal coefficients and detail how we use it for the 3d Ising CFT in section~\ref{sec:details}.

\subsection{The 3d Ising CFT}
\label{sec:3dIsingReview}

In this work, we apply the thermal crossing equation and inversion formula to compute thermal one-point coefficients in the 3d Ising CFT. It will be crucial to incorporate as much information as possible about the known flat-space data (i.e.\ operator dimensions and OPE coefficients) of the theory. Indeed, our approach will be closely tailored to observed features of this data. We leave the question of how our approach can be generalized to arbitrary CFTs for future work. In this section, we review some features of the spectrum of the 3d Ising CFT that play an important role in what follows.

\begin{table}
\begin{center}
{\small
\begin{tabular}{|c|c|c|l|l|l|l|l|}
\hline
$\cO$ & family & $\Z_2$ & $\ell$ & $\De$ & $\tau=\De-\ell$ & $f_{\s\s\cO}$ & $f_{\e\e\cO}$ \\
\hline
$\e$ & ? & $+$ & 0 & $1.412625{\bf\boldsymbol(10\boldsymbol)}$ & $1.412625{\bf\boldsymbol(10\boldsymbol)}$ & $1.0518537{\bf\boldsymbol(41\boldsymbol)}$ & $1.532435{\bf\boldsymbol(19\boldsymbol)}$ \\
$\e'$ & $[\s\s]_1$ & $+$ & 0 & $3.82968(23)$ & $3.82968(23)$ & $0.053012(55)$ & $1.5360(16)$ \\
$T_{\mu\nu}$ & $[\s\s]_0$ & $+$ & 2 & $3$ & $1$ & $0.32613776(45)$ & $0.8891471(40)$ \\
$T'_{\mu\nu}$ & $[\s\s]_1$ & $+$ & 2 & $5.50915(44)$ & $3.50915(44)$ & $0.0105745(42)$ & $0.69023(49)$ \\
$C_{\mu\nu\rho\s}$ & $[\s\s]_0$ & $+$ & 4 & $5.022665(28)$ & $1.022665(28)$ & $0.069076(43)$ & $0.24792(20)$ \\
\hline
\hline
$\cO$ & family & $\Z_2$ & $\ell$ & $\Delta$ & $\tau=\De-\ell$ & $f_{\s\e\cO}$ & -  \\
\hline
$\s$ & ? & $-$ & 0 & $0.5181489{\bf\boldsymbol(10\boldsymbol)}$ & $0.5181489{\bf\boldsymbol(10\boldsymbol)}$ & $1.0518537{\bf\boldsymbol(41\boldsymbol)}$ & \\
$\s'$ & ? & $-$ & 0 & $5.2906(11)$ & $5.2906(11)$ & $0.057235(20)$  &\\
&  $[\s\e]_0$&  $-$ & 2 & $4.180305(18)$ & $2.180305(18)$ & $0.38915941(81)$ & \\
\hline
\end{tabular}
}
\end{center}
\caption{
A few low-dimension operators in the 3d Ising CFT, from \cite{Simmons-Duffin:2016wlq}. The ``?'' are associated to scalars whose affiliation with a certain operator family is not fully established. Errors in bold are rigorous. All other errors are non-rigorous. 
}
\label{tab:lowestdim}
\end{table}

The low-dimension spectrum of the 3d Ising CFT is summarized in table~\ref{tab:lowestdim}. The lowest-dimension operator is a $\Z_2$-odd scalar $\s$ with dimension $\De_\s\approx 0.518$. The lowest-dimension $\Z_2$-even scalar $\e$ has dimension $\De_\e\approx 1.412$.

Some of the operators in table~\ref{tab:lowestdim} are (conjecturally) identifiable as members of large-spin families --- i.e.\ families of operators whose twists $\tau=\De-\ell$ accumulate at large spin. This identification works as follows. At asymptotically large spin, it is known that there exist ``multi-twist" operators $[\cO_1\cdots\cO_k]_{n,\ell}$ whose twists approach $\tau_1+\cdots +\tau_k+2n$ as $\ell\to \oo$, where $\tau_i=\De_{\cO_i}-\ell_{\cO_i}$ \cite{Simmons-Duffin:2016wlq}.  By analyticity in spin, all operators $\cO$ with spin above the Regge intercept $\ell>\ell_0$ are expected to lie on curves $\tau_i(\ell)$ that are analytic in $\ell$ \cite{Caron-Huot:2017vep, Simmons-Duffin:2017nub}. Here, $i$ labels the Regge trajectory of the operator. If the trajectory associated to $\cO$ approaches a multi-twist value $\tau_1+\cdots+\tau_k+2n$ as $\ell\to \oo$, we say that $\cO$ is in the family $[\cO_1\cdots\cO_k]_{n}$. In practice, to identify a particular family in numerics, one computes Regge trajectories using the lightcone bootstrap \cite{Simmons-Duffin:2016wlq,Fitzpatrick:2012yx,Komargodski:2012ek,Alday:2015eya,Alday:2015ota,Alday:2015ewa,Alday:2016njk} or Lorentzian inversion formula \cite{Caron-Huot:2017vep, Simmons-Duffin:2017nub,Kravchuk:2018htv} and observes which operators they pass through.\footnote{Operator mixing can make this procedure difficult in practice. Due to eigenvalue repulsion it may be difficult to track a trajectory out to infinite spin if it passes near other trajectories. It is also not known rigorously whether trajectories remain discrete in twist space when $\ell$ is not an integer. See \cite{Simmons-Duffin:2016wlq} for further discussion. None of these subtleties are visible in the first few orders of large-spin perturbation theory.}${}^,$\footnote{The operators marked with ``?" in table~\ref{tab:lowestdim} are scalars. Whether scalar operators lie on Regge trajectories depends on the behavior of four-point functions in the Regge regime. It has been conjectured that scalars do lie on Regge trajectories in the 3d Ising CFT \cite{SimonTalk}.}
\begin{figure}[t!]
\begin{center}
\includegraphics[width=0.7\textwidth]{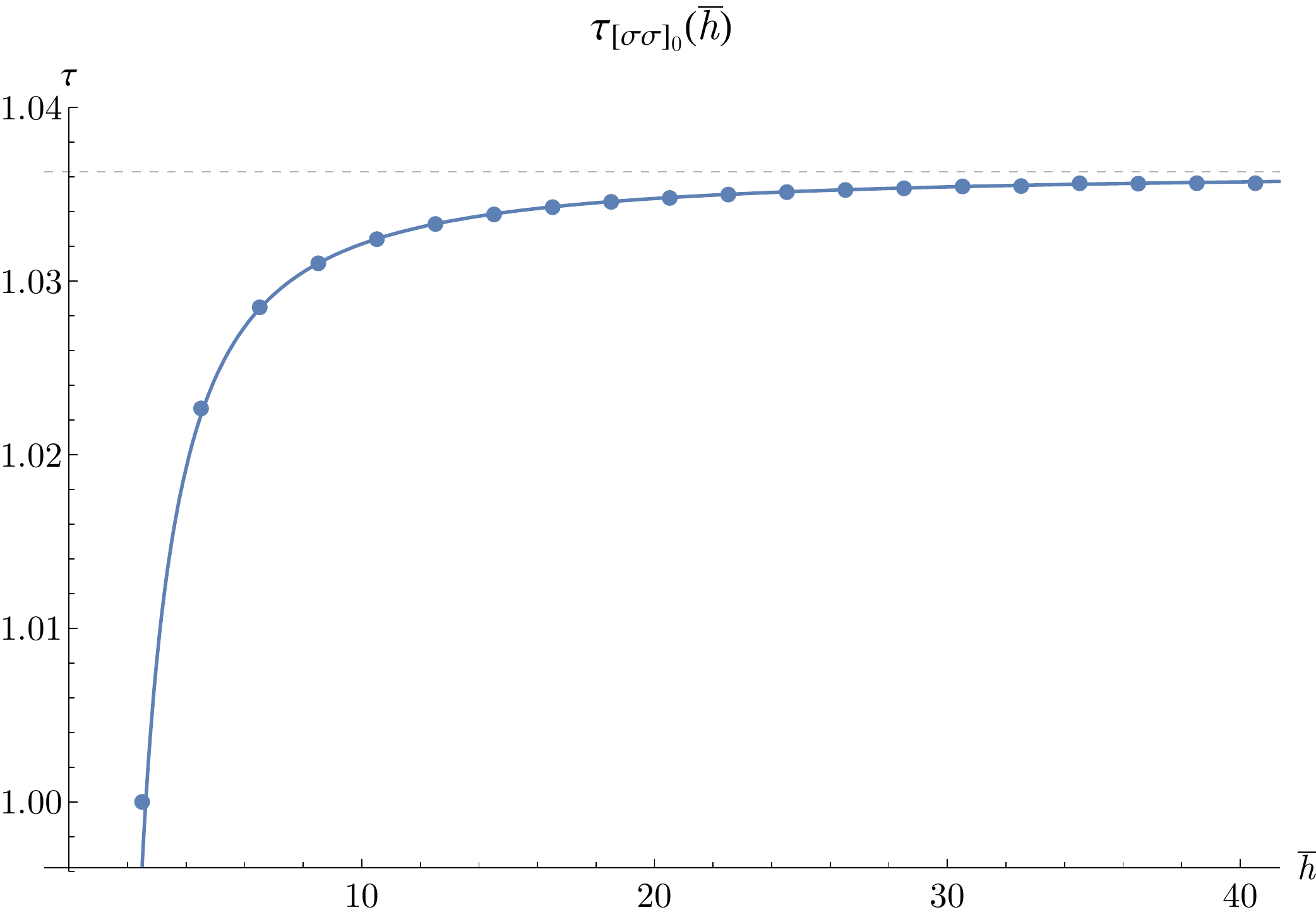}
\end{center}
\caption{Twists of the double-twist family $[\s\s]_0$. Here, we plot $\tau=\De-\ell$ versus $\bar h = \frac{\De+\ell}{2}$. The dots show estimates from \cite{Simmons-Duffin:2016wlq} using the extremal functional method \cite{Poland:2010wg,ElShowk:2012hu,El-Showk:2014dwa} and the numerical bootstrap. The curve shows the prediction of large-spin perturbation theory with only $\De_\s,\De_\e,f_{\s\s\e},c_T$ taken from the numerical bootstrap. figure reproduced from \cite{Simmons-Duffin:2016wlq}.}
\label{fig:firstfamily}
\end{figure}

Numerical bootstrap methods reveal large OPE coefficients in the $\s\x\s$ and $\e\x\e$ OPEs for operators in the families $[\s\s]_0$,  $[\e\e]_0$, and $[\s\s]_1$. Certain other trajectories with comparable twist are not described to high precision by numerics, including for example $[\s\s\s\s]_0$. Instead, numerics indicate that these other families have relatively small OPE coefficients in the $\s\x\s$ and $\e\x\e$ OPEs. In this work, we make the approximation that we can ignore large-spin families other than $[\s\s]_0$, $[\e\e]_0$, and $[\s\s]_1$. It is difficult to quantify the error associated with this approximation, since other families could potentially possess large thermal one-point coefficients that don't play a role in flat-space correlators, but do contribute to thermal correlators. Nevertheless, we will find a mostly-consistent picture. However, we also see some indications that other families (in particular $[\s\s\e]_0$) could be important for more precise calculations, see section~\ref{section:mixing between families}.

Let us discuss the families $[\s\s]_0$, $[\e\e]_0$, and $[\s\s]_1$ in more detail. The lowest-twist family $[\s\s]_0$ has twists ranging from $1$ at $\ell=2$ to $2\De_\s = 1.036$ as $\ell\to \oo$. They are increasing and concave-down as a function of $\ell$, by Nachtmann's theorem \cite{Nachtmann:1973mr,Komargodski:2012ek,Costa:2017twz}. The lowest-spin operator in the $[\s\s]_0$ family is the spin-2 stress-tensor $T_{\mu\nu}$. The next operator $C_{\mu\nu\rho\s}$ has spin-4 and controls the breaking of cubic symmetry when the Ising model is implemented on a cubic lattice \cite{Campostrini:1999at}. The family $[\s\s]_0$ is plotted up to spin 40 in figure~\ref{fig:firstfamily}. There we show both the numerical bootstrap predictions (dots) and the results of large-spin perturbation theory (curve), which agree to high precision \cite{Alday:2015ewa,Simmons-Duffin:2016wlq}. The curve $\tau_{[\s\s]_0}(\bar h)$ is well-approximated by $2(2h_\s+ \delta_{[\s\s]_0}(\hb))$ where $h_\s=\De_\s/2$ and 
\be
\label{eq:delta ss0}
\delta_{[\s\s]_0}(\hb) &= \frac{\sum _{\cO=\e,T} -f_{\sigma \sigma \cO}^2\frac{ \Gamma (2
   \bar{h}_{\cO})}{\Gamma (\bar{h}_{\cO}){}^2}  Q_{h_{\cO}-\Delta _{\sigma
   }}(\bar{h})}{Q_{-\Delta _{\sigma }}(\bar{h})-\sum
   _{\cO=\e,T} 2 f_{\sigma \sigma \cO}^2 \left(\psi ^{(0)}(\bar{h}_{\cO})+\gamma
   \right) \frac{ \Gamma (2
   \bar{h}_{\cO}) }{\Gamma
   (\bar{h}_{\cO}){}^2} Q_{h_{\cO}-\Delta _{\sigma }}(\bar{h}) }\nn  
\\ &=\frac{-0.000971264\frac{ \Gamma (\hb-0.981851)}{\Gamma (\hb+0.981851)}-0.031588\frac{ \Gamma (\hb-1.18816)}{\Gamma (\hb+1.18816)}}{0.68256\frac{ \Gamma (\hb-0.481851)}{\Gamma (\hb+0.481851)}-0.00248716\frac{ \Gamma (\hb-0.981851)}{\Gamma (\hb+0.981851)}+0.0394879\frac{ \Gamma (\hb-1.18816)}{\Gamma (\hb+1.18816)}}\, ,
\ee with $Q_a(\hb) = \frac{1}{\Gamma(-a)^2}\frac{\Gamma(\hb-a-1)}{\Gamma(\hb+a+1)}$. The OPE coefficients of $[\s\s]_0$ in the $\s\x\s$ and $\e\x\e$ OPEs can also be approximated in large-spin perturbation theory and are given in \cite{Simmons-Duffin:2016wlq}.

\begin{figure}[t!]
\begin{center}
\includegraphics[width=0.7\textwidth]{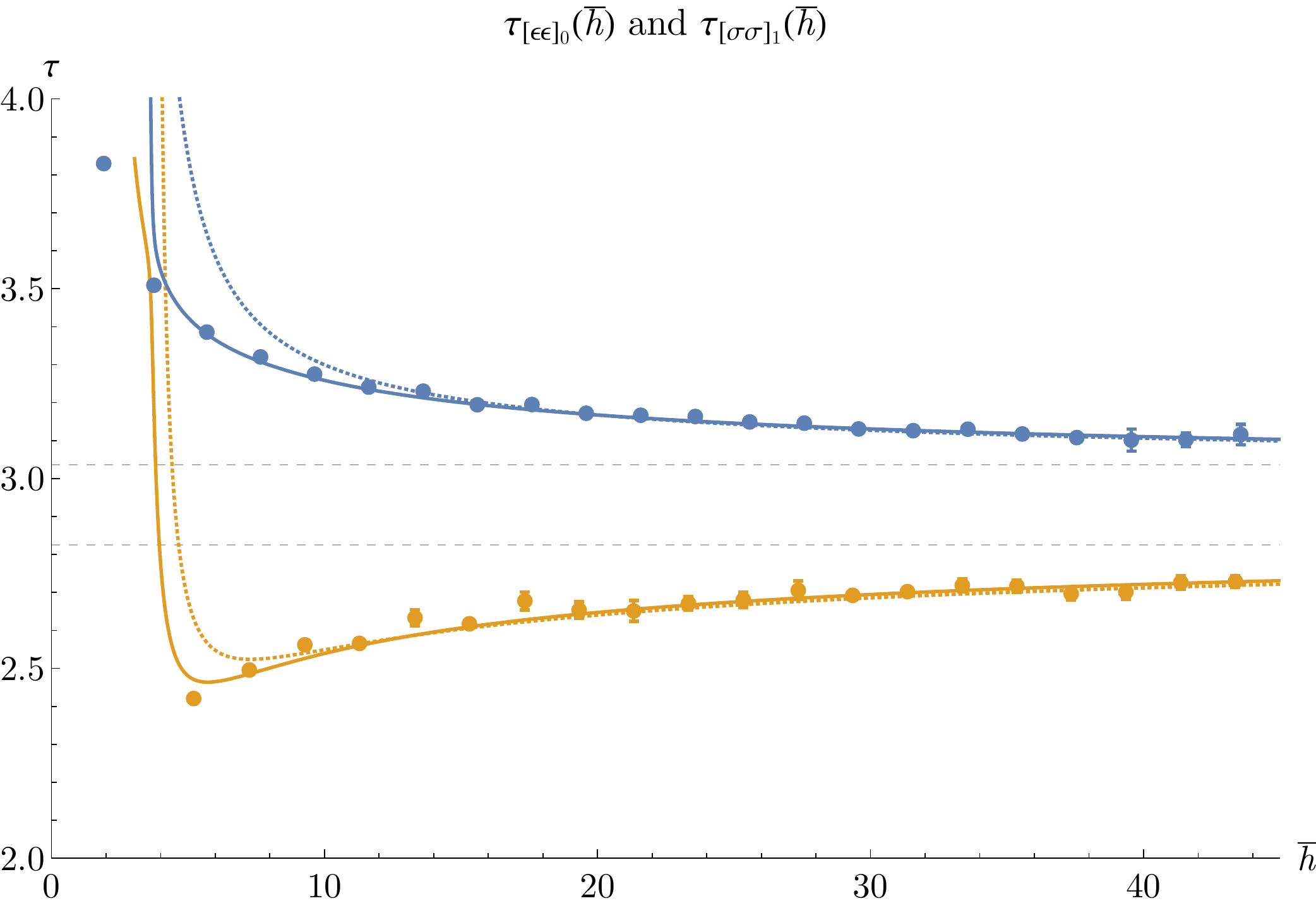}
\end{center}
\caption{Twists of the double-twist families $[\e\e]_0$ (orange) and $[\s\s]_1$ (blue). Again, we plot $\tau=\De-\ell$ versus $\bar h = \frac{\De+\ell}{2}$. The dots show estimates using the extremal functional method and the numerical bootstrap. The curves are estimates using large-spin perturbation theory and the mixing procedure described in \cite{Simmons-Duffin:2016wlq} and reviewed in section~\ref{section:mixing between families}. The dashed curves illustrate the effects of modifying the mixing procedure. Figure reproduced from \cite{Simmons-Duffin:2016wlq}.}
\label{fig:nexttwofamilies}
\end{figure}

The families $[\e\e]_0$ and $[\s\s]_1$ are notable in that they experience large mixing with each other at small spins. For example, the operators $[\s\s]_1$ have larger OPE coefficients than $[\e\e]_0$ in the $\e\x\e$ OPE for spins $\ell\lesssim 25$. This mixing can be described by supplementing large-spin perturbation theory with a procedure described in \cite{Simmons-Duffin:2016wlq}. The resulting twists and OPE coefficients match well with estimates using the extremal functional method which is used in the numerical bootstrap to extract the spectrum of theories on the boundary of the allowed region \cite{El-Showk:2014dwa}. We show the twists of the $[\s\s]_1$ and $[\e\e]_0$ families in figure~\ref{fig:nexttwofamilies}.

\section{Method and results}
\label{sec:method-and-results}

\subsection{Summary of method}
\label{sec:method-summary}

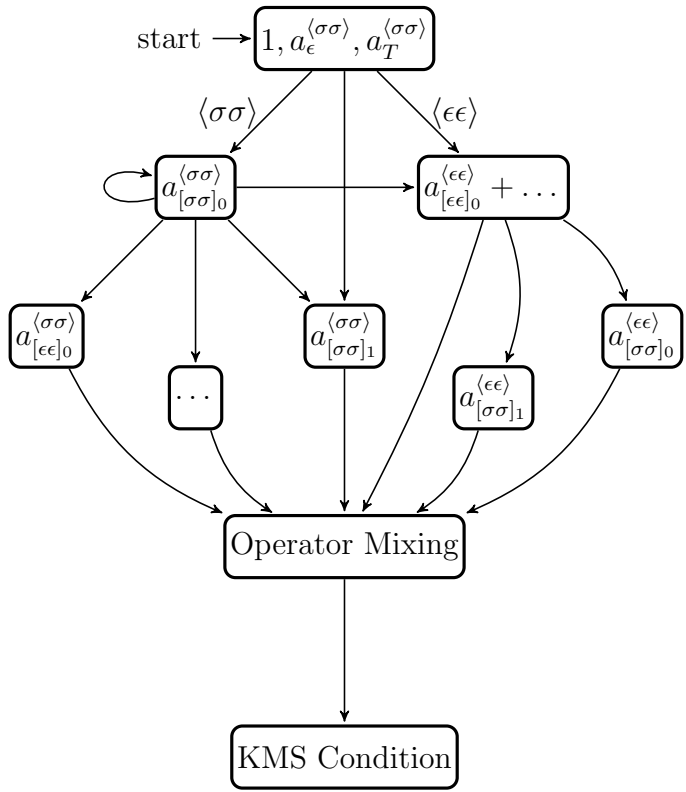
\begin{figure}[t!]
\begin{center}
\begin{tikzpicture}[->,>=stealth',shorten >=1pt,auto,node distance=2.8cm,
                    semithick]
  \tikzstyle{every state}=[fill=red,draw=none,text=white]

  \node[state] (A)                    {$a_{[\s\s]_0}^{\<\s\s\>}$};
  \node[initial, state]         (B) [above right of=A] {$1, a_\e^{\<\s\s\>}, a_T^{\<\s\s\>}$}; 
    \path (B)
            edge              node[left] {$\langle\s\s \rangle$} (A); 
     \path (A)  	edge[loop left]    node[anchor=north,above]{} (A);       
         \node[state]         (D) [below left of=A] {$a_{[\e\e]_0}^{\<\s\s\>}$};
        \node[state]         (E) [below right of=A] {$a_{[\s\s]_1}^{\<\s\s\>}$};
     \node[state]         (F) [below of=A] {$\dots$};
     \path   (A)          edge              node {} (D)
            edge             node {} (E)
             edge             node {} (F)
             (B) edge node{} (E);
  \node[state]         (C) [below right of=B] {$a_{[\e\e]_0}^{\<\e\e\>}+\dots$};
 \path   (B)         edge             node[right] {$\langle\e\e \rangle$} (C)
             (A) edge node{} (C);
             
  \node[state]         (H) [below of=C] {$a_{[\s\s]_1}^{\<\e\e\>}$};
  
  \node[state]         (G) [below right of=C] {$a_{[\s\s]_0}^{\<\e\e\>}$};
  
  \path      
            (C)
                	edge[bend right=-20]  node[anchor=south,above]  {} (H) 
(C) edge[bend left=20]              node {} (G);                
   \node[state]         (I) [below of=E] {Operator Mixing};
    \path            	(D) edge[bend right=20]              node {} (I)
                			(F) edge[bend right=20]              node {} (I)
                				(E) edge[bend left=0]              node {} (I)

                				(G) edge[bend left=20]              node[anchor=north,above] {} (I)
                				(H) edge[bend left =20]              node[anchor=north,above] {} (I)
                				(C) edge[bend left = 5 ]              node {} (I) ;                             
				   
      \node[state]         (J) [below of= I] {KMS Condition};
                	\path		    (I) edge[bend right = 0 ]              node {} (J);
\end{tikzpicture}
\end{center}
\caption{Diagram of the algorithm which is used to obtain the thermal coefficients in the 3d Ising CFT. Here, ``...'' represents contributions to the thermal coefficients of families other than $[\s\s]_0$, $[\s\s]_1$, and $[\e\e]_0$ that we account for when considering operator mixing. }
\label{fig:flow-chart-thermal-coeff}
\end{figure}

The thermal bootstrap for the 3d Ising CFT consists of two parts. In the first part, we compute the thermal coefficients of a truncated (but infinite) subset of the spectrum in terms of the thermal coefficients of a few operators --- $\mathbf{1}$, $\e$, and $T$, where $a^{\<\s\s\>}_\e$ and $a^{\<\s\s\>}_T$ are unknowns. Specifically, we use the thermal inversion formula to approximately determine the thermal coefficients of all operators in the $[\s\s]_0$, $[\s\s]_1$, and $[\e\e]_0$ families described in section~\ref{sec:3dIsingReview}. 
In the second part, we approximate $\<\s\s\>$ 
as a sum over the truncated spectrum with the thermal coefficients obtained in the first part. We determine the remaining unknowns by demanding that the KMS condition is satisfied in a region of the $(z, \bar z)$-plane that is within the radius of converge of the s-channel OPE. The procedure is summarized graphically in figure \ref{fig:flow-chart-thermal-coeff}. The initial steps are as follows:
\begin{enumerate}
\item Consider the thermal inversion formula for the $\<\s\s\>$ correlator. 
\item Invert the low-twist operators $\mathbf{1}$, $\e$, and $T$ in the $t$-channel OPE to compute $a_{[\s\s]_0}^{\<\s\s\>}(J)$ in terms of the unknowns $a_{\e,T}^{\<\s\s\>}$.
\item Sum over the $[\s\s]_0$ family in the $t$-channel using the computed data. Invert the result to obtain self-corrections of the $[\s\s]_0$ family.
\item  
Compute poles for higher-twist families up to twist 2 in the $\<\s\s\>$ and $\<\e\e\>$ correlators by summing the self-corrected thermal coefficients of the $[\s\s]_0$ family together with $\mathbf{1}$, $\e$, and $T$.
\item Estimate the thermal coefficients of the $[\s\s]_1$ and $[\e\e]_0$ families at intermediate spin by ``mixing'' the residues according to the large anomalous dimensions. 
\item Assuming the smoothness of the thermal coefficients in the $[\s\s]_1$ family with $\bar h$ up to $J=0$, we interpolate the thermal coefficients of the $[\s\s]_1$ family to estimate the thermal coefficients of $\e'$ and $T'$.
\item After these steps, we are almost ready to determine the unknowns. As a penultimate simplification, we use the fact that $T$ is the spin-two member of the $[\s\s]_0$ family. 
This requires that $a_T^{\<\s\s\>}$ is equal to $a_{[\s\s]_0}^{\<\s\s\>}(J=2)$, which we use to solve for $a_T^{\<\s\s\>}$. Thus, we are left with a single unknown, $a_{\e}^{\<\s\s\>}$. 
\end{enumerate}

Finally, we approximate the $\<\s\s\>$ correlator by the truncated OPE including the scalars $\mathbf{1}$ and $\e$, and the low-twist families $[\s\s]_0$, $[\s\s]_1$, and $[\e\e]_0$. We solve for the final remaining unknown $a_\e^{\<\s\s\>}$ by imposing that the KMS condition is close to being satisfied for a sampling of $z$ and $\bar z$ points in the interior of the square $0\le z,\bar z\le 1$.

\subsection{Results}
\label{sec:results}

In this section, before diving into the details of our computation, we summarize our results and compare to MC.  To perform our computation, we must make some arbitrary choices and approximations. We enumerate them in section~\ref{sec:error-bound} and estimate the resulting errors. Overall, the results show robustness for a wide range of choices.

\subsubsection{One-point functions}
\label{sec:one-point-func-results}

After using the thermal inversion formula together with the KMS condition we find that 
\be
\label{eq:final-results}
a_\e^{\<\s\s\>} = 0.672(74), \qquad a_T^{\<\s\s\>} =1.96(2)\,, \qquad b_\epsilon = 0.63(7) \,, \qquad b_T = -0.43(1).
\ee 
The values and errors quoted capture the deviations seen  over several runs of our algorithm with different parameter choices. For comparison the results obtained from MC are 
\be
\label{eq:MC-results}
b_{\epsilon}^{\text{MC}}= 0.667(3)\,\text{\cite{HasenbuschPrivate}}\,,\qquad \qquad b_{T}^{\text{MC}}= -0.459(3)\,\text{\cite{PhysRevE.79.041142,PhysRevE.53.4414,PhysRevE.56.1642}}\,.
\ee
Note that the errors for the above two observables in MC are much smaller than for the bootstrap. This is due in part to the difficulty of using the thermal crossing equation, and also to the favorable behavior of finite-size effects when computing thermal correlators with MC, see appendix~\ref{app:mc}. Improving the precision of thermal bootstrap results is clearly an important challenge for the future.

\begin{figure}[t!]
\centering{
\includegraphics[width=0.75\textwidth]{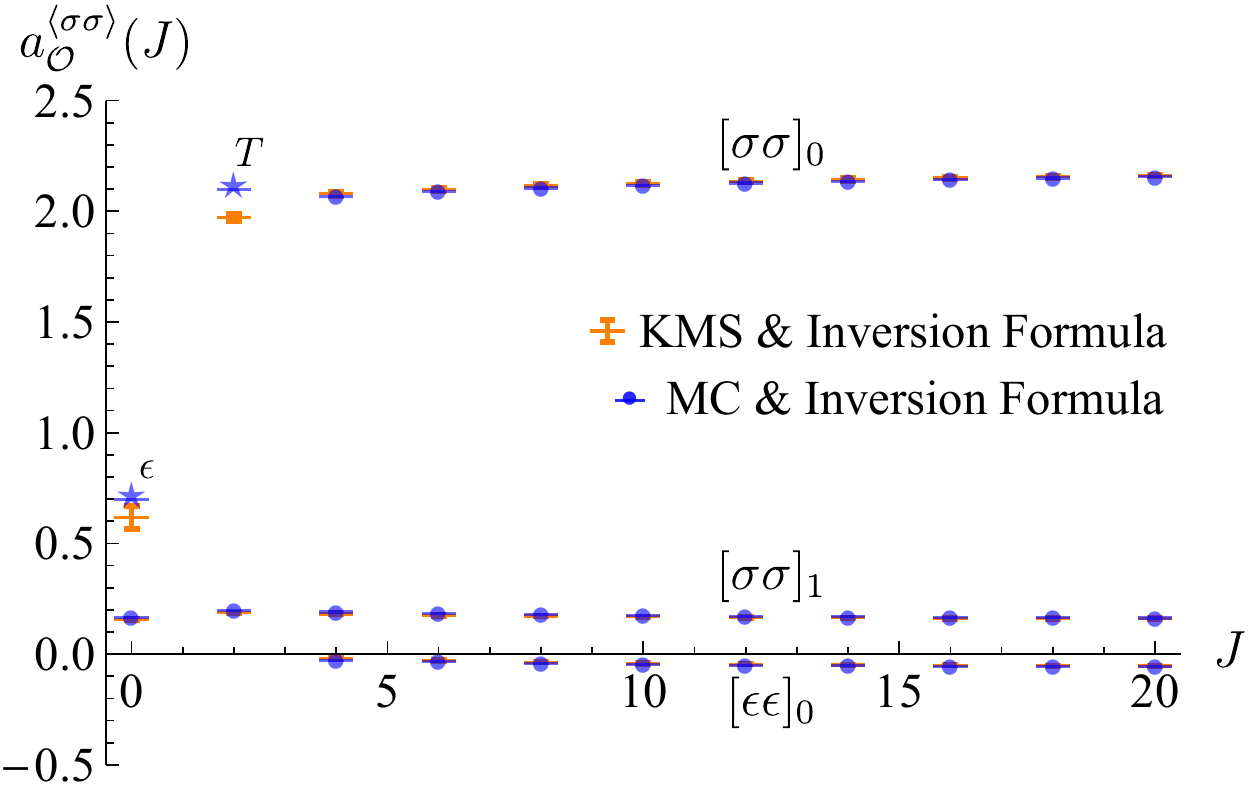}}
\caption{Thermal coefficients for the three families $[\s\s]_0$, $[\s\s]_{1}$, and $[\e\e]_0$. The orange horizontal lines are obtained by using the KMS condition in combination with the thermal inversion formula and by averaging over several parameter choices. The spread given by the orange error bars is obtained by computing the operator mixing using different sets of $\bar z$ values as explained in Section \ref{section:mixing between families} and by imposing the KMS conditions in different regions on the thermal cylinder (see figure \ref{fig:zzbar region for KMS} for an example). The blue stars are MC estimates for  $a_\e^{\<\s\s\>,\text{~MC}} = 0.711(3)$  \cite{HasenbuschPrivate} and $a_T^{\<\s\s\>,\text{~MC}} = 2.092(13)$ \cite{PhysRevE.79.041142,PhysRevE.53.4414,PhysRevE.56.1642}. The blue lines are the estimates for the thermal coefficients of all other operators in $[\s\s]_0$, $[\s\s]_{1}$ and $[\e\e]_0$ families using these MC results together with the inversion formula. Note that the spread of the thermal coefficients of higher-spin operators estimated by the bootstrap are too small to be visible on this scale. }
\label{fig:thermal coefficients of families}
\end{figure}

Our determinations for thermal coefficients in the three low-twist families, $[\s\s]_0$, $[\s\s]_1$ and $[\e\e]_0$ are presented in figure \ref{fig:thermal coefficients of families}.\footnote{We choose to present the thermal coefficients $a_\cO^{\<\s\s\>}$ instead of the thermal one-point function due to the exponential increase of $b_\cO$ with the spin $J$ (see definition \eqref{eq:two-pt-function-expansion}). } Unfortunately, to our knowledge, there are no available MC results for the thermal one-point functions of such higher-spin operators. However, we can use the MC results for $\epsilon$ and $T$ in \eqref{eq:MC-results} together with the thermal inversion formula to compare to the results obtained in our computation.  Note that due to the strong contribution of the unit operator in the inversion formula, the standard deviations in the thermal coefficient of all higher-spin operators in all three families is much smaller than that for $a_\e^{\<\s\s\>} $ and $a_T^{\<\s\s\>}$. 

\subsubsection{Two-point function of $\s$}
\label{sec:two-point-func-results}

\begin{figure}[t!]
\centering{
\includegraphics[width=0.49\textwidth]{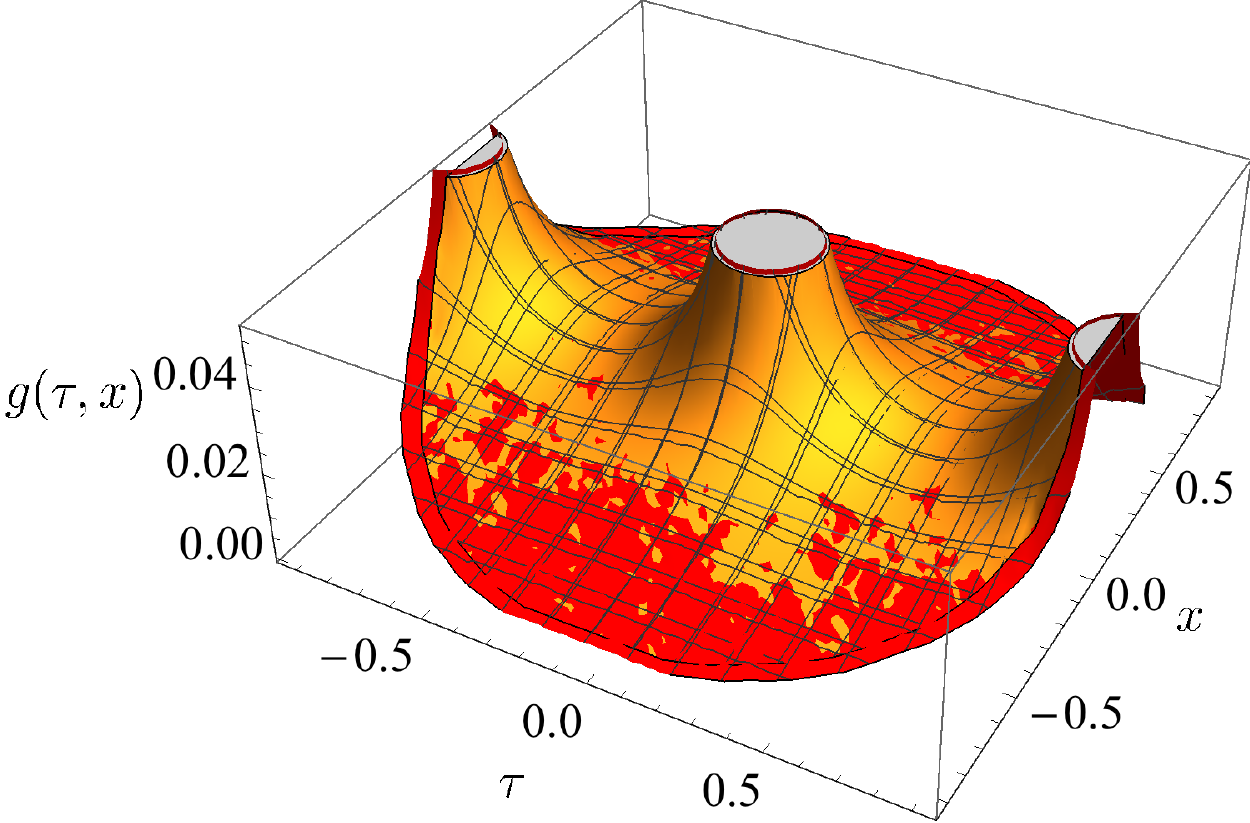}
\includegraphics[width=0.49\textwidth]{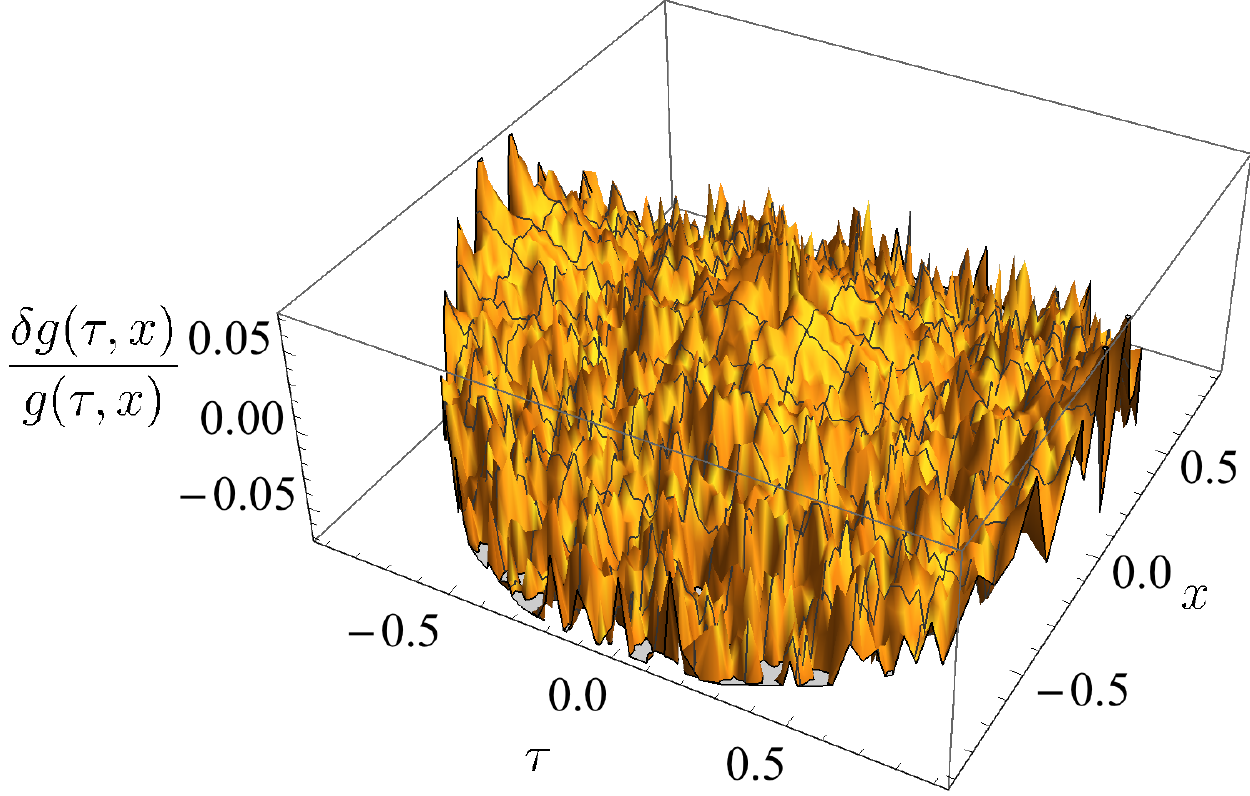}
}
\caption{\textit{Left:\/} The thermal two-point function obtained by applying the inversion formula and then solving the KMS condition (yellow) compared to that obtained from a MC simulation (red). Note that we restrict the plot to the region of OPE convergence around $x = 0$ and $\tau = 0$. \textit{Right:\/} Percentage difference between the two correlators, showing good agreement (within $~5\%$) between the bootstrap and MC predictions. At small values of $\sqrt{|x|^2 + \tau^2}$ we expect the MC results to be inaccurate due to lattice-size effects. As $\sqrt{|x|^2 + \tau^2} \rightarrow \beta$, we exit the region of OPE convergence, and we expect inaccuracies in the bootstrap calculation.  }
\label{fig:2pt function plots}
\end{figure}

In figure~\ref{fig:2pt function plots}, we show the thermal two-point function $\<\s\s\>_\b$ computed using our algorithm and compare it to a MC simulation that we performed. The details of our simulation are described in appendix~\ref{app:mc}.

Overall, we find good agreement between the bootstrap prediction and MC inside the regime of convergence of the OPE. In part, this is due to the fact that the unit operator gives a large contribution in this region, and its contribution is known very precisely from the four-point function bootstrap. However, the thermal OPE also correctly recovers other features of the two-point function. For example, large-spin families sum up to correctly reproduce the $t$-channel singularity as $\tau\to \pm 1$.

We also observe decay of the two-point function in the spatial direction $x$. Exponential decay of thermal two-point functions in $x$ can established rigorously by expanding the correlator in states on $\R\x S^1$, as explained in \cite{Iliesiu:2018fao,Bros:1995he}. However, decay in $x$ is not obvious from the OPE, where each term grows in magnitude in the $x$ direction. The fact that we observe decay in $x$ serves as a check on our calculation. At long distances, the correlator behaves as $e^{-m_\mathrm{th} x}$, where $m_\mathrm{th}$ is the thermal mass. It would be interesting to understand how to determine or bound $m_\mathrm{th}$ using information in the OPE region.\footnote{We thank Tom Hartman for discussions on this point.}

\begin{figure}[tb]
\hspace{1.0cm}
\includegraphics[width=.75\textwidth]{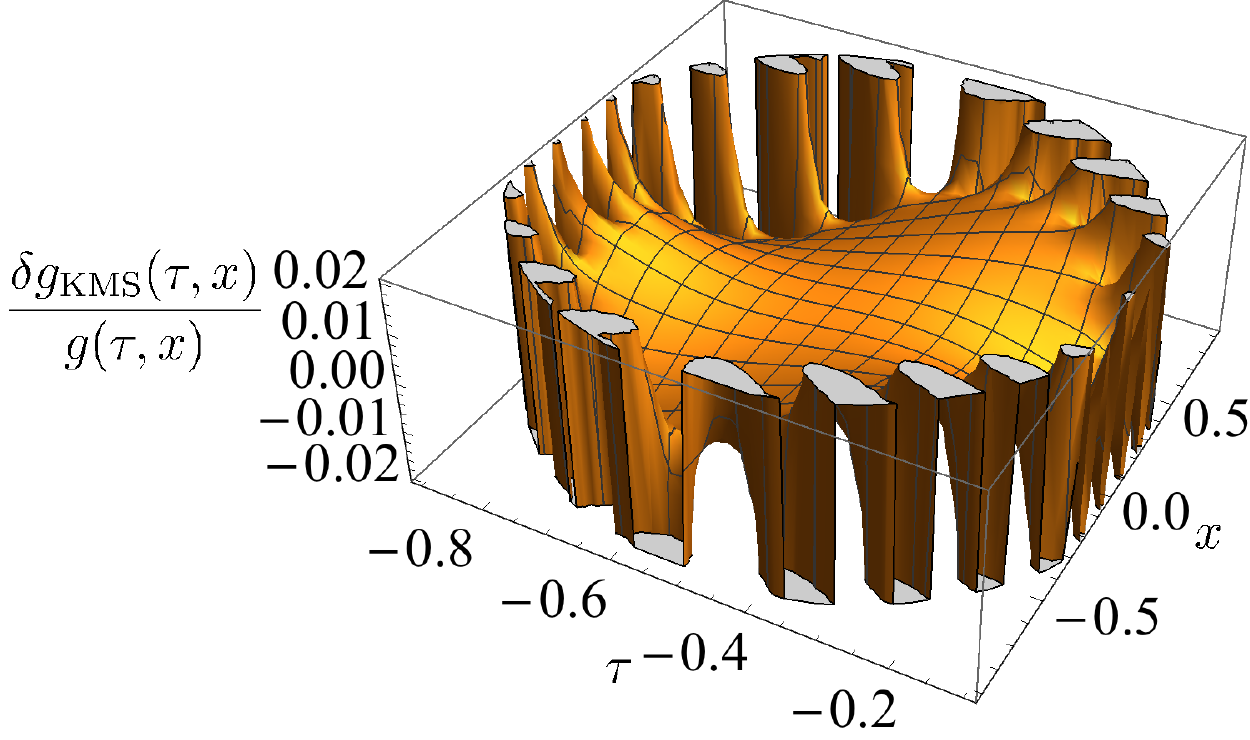}
\label{fig:symm-region-KMS}
\caption{Evidence of how well the KMS condition is satisfied in the $( \tau,{x})$ plane in the region of OPE convergence. We plot the difference of the two-point function and it's periodic image, $\delta g_{\text{KMS}}(\tau, x) = g(1+\tau,x) - g(\tau,x)$, using the average thermal coefficients presented in figure \ref{fig:thermal coefficients of families}. Note that towards the boundary of the region of OPE convergence, our estimates for the two-point function become worse and the KMS condition is further from being satisfied. For the range $(\tau, x)$ shown above the deviation from satisfying KMS is $<2\%$.    }
\label{fig:diff-crossing}
\end{figure}

Finally, in figure \ref{fig:diff-crossing} we test how close we are to satisfying the KMS condition within the region of OPE convergence.   As emphasized in the figure, within an $0.9 \beta$ radius from the center of the OPE convergence region the deviation from satisfying KMS is $<2\%$.

\subsubsection{Systematic errors}
\label{sec:error-bound}

Our algorithm above involves a few choices of parameters. To check for robustness under different choices, we show the spread of results for the thermal coefficients in figure \ref{fig:thermal coefficients of families}. Specifically, variations in our results are mainly due to the following choices:    
\begin{itemize}
\item As we explain in section~\ref{section:mixing between families}, the mixing of families requires a set of $\zb$ points. Figure \ref{fig:thermal coefficients of families} shows the results obtained when choosing different sets of $\zb$ values which span a full order of magnitude. When considering our results for $a_\e^{\< \s\s\>}$, the variation between the set with the lowest values of $\zb$ and those with the largest is at most $\sim 10\%$. As we will describe in section~\ref{section:mixing between families} and is already clear from figure \ref{fig:thermal coefficients of families}, the error for higher spin thermal coefficients is significantly lower. 
\item In the final step of our algorithm, we choose a set of point in the $(z,\zb)$-plane, in the $s$-channel region of convergence, for which we require that the thermal two-point function satisfies the KMS condition approximately. When considering significantly different regions in the $(z,\zb)$-plane as exemplified in figure \ref{fig:zzbar region for KMS}, the variation in $a_\e^{\< \s\s\>}$ is only $\sim 5\%$. Once again, the error associated to this effect for operators with higher spins is significantly lower. 
\end{itemize}
Besides the choices of parameters presented above, there are several other systematic errors:
\begin{itemize}
\item When requiring that the KMS condition is close to being satisfied at a wide variety of point in the $(z, \bar z)$-plane, we truncate the OPE of the two-point function $\<\s\s\>$ to the three low-twist families  $\<\s\s\>$.  For the ranges of points at which we attempt to impose the KMS condition, corrections to the two-point function are dominated by the contribution of the next $\mathbb Z_2$-even operator $\epsilon''$.\footnote{We remind the reader that this operator is not part of any of the three double-twist families  $[\s\s]_0$, $[\s\s]_1$ and $[\e\e]_0$. } Considering that the flat-space numerical bootstrap estimates the  scaling dimension of this operator to be $\Delta_{\epsilon''} \sim 6.9 $, we can compare the contribution of the thermal conformal block for this operator to the total contribution of all other operators in $[\s\s]_0$, $[\s\s]_1$, and $[\e\e]_0$ to $\<\s\s\>$. This helps us estimate the error associated with neglecting this operator and higher twist operators to be $\sim 4\%$.  

\item The second largest systematic error which we expect comes from the fact that when using the inversion formula we truncate the range of integration to the t-channel region of convergence, $z\le 2$. As discussed in section~\ref{sec:thermal-review} we expect that the correction from the region $z \geq 2$ to the thermal coefficient of an operator with spin $J$  is exponentially suppressed in $J$. However, since we use the inversion formula for $J \geq 4$, one might worry that at small $J$ this correction becomes large. To probe this we note that in the $O(N)$-model with $N \rightarrow \infty$ the difference between the exact result and that extracted by inverting the OPE for an operator with $ J=4$ is only $\sim 2.8\%$.\footnote{Specifically, the exact result found in \cite{Iliesiu:2018fao} predicts that as $N \rightarrow \infty$, $a_{[\s\s]_{0, \ell = 4}}^{\text{exact},\,\<\s\s\>} = 0.964$, while by restricting the inversion formula to the interval $1 \leq z \leq 2$,  we find  $a_{[\s\s]_{0, \ell =4}}^{\text{OPE}\,\<\s\s\>} = 0.936$. }

\item  There are several systematic errors associated to the operator mixing procedure. The first is due to the truncation of the spectrum to operators of twist below a cut-off value.   Since the contribution of operators with higher twist is visibly suppressed, such a truncation should only introduce a small error. The second is due to the fact that while multiple operator families serve as mixing inputs, we solely focus on $[\s\s]_0$, $[\s\s]_1$, and $[\e\e]_0$ as outputs. This assumes that, just like in the flat-space bootstrap, the thermal coefficients of these three double-twist families dominate over all other families with twists below the cut-off. While we have found this to be true for the thermal coefficients in the $\<\s\s\>$ correlator, there is one family --- the multi-twist family $[\s\s\e]_0$ --- which has a contribution comparable to that of $[\e\e]_0$ in the $\< \e\e\>$ correlator. While we will discuss the contribution of this family extensively in section~\ref{sec:half-invert}, here we note that neglecting its contribution in the mixing procedure leads to an overall difference of $\sim 4\%$ in the mixing results. Finally, we note that after mixing we assume that the $[\s\s]_1$ family is smooth in $\bar h$ and we use a fit to estimate the thermal coefficients of the $\e'$ and $T'$ operators. We find that by varying this fit we introduce an overall error of $\sim 3\%$ in the final results.

\end{itemize}

\section{Details of the computation}
\label{sec:details}

In this section, we describe the details of the algorithm outlined in section~\ref{sec:method-summary}. We will methodically iterate large-spin perturbation theory --- working our way up in twist --- to compute the thermal coefficients for the $[\s\s]_0$, $[\s\s]_1$, and $[\e\e]_0$ families.

In general, we will invert operators with $h<1$ from the $t$-channel, meaning we will work to order 
\begin{align}
S_{1-2h_\s,2h_\s}(\hb) \sim \frac{1}{\hb^{2-2h_\s}}
\end{align} for the thermal coefficients in the $\<\s\s\>$ correlator, dropping terms $S_{c,\Delta}(\hb)$ with $c> 1-2h_\s$, and analogously for $\<\e\e\>$ with $h_\s$ replaced with $h_\e$.

\subsection{$[\s\s]_0$} 
\label{sec:ss0}

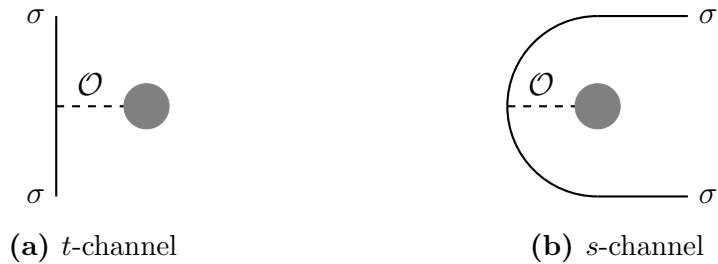
\begin{figure}[tb]
\centering
\begin{subfigure}[t]{.4\textwidth}
\centering
\begin{tikzpicture}[xscale=0.6,yscale=0.6]
\draw[thick] (0,0) -- (0,4);
\draw[thick, dashed] (0,2) -- (1.5,2);
\node[left] at (0,0) {$\s$};
\node[left] at (0,4) {$\s$};
\node[above] at (.75,2) {$\cO$};
\filldraw[gray] (2,2) circle (.5);
\end{tikzpicture}
\caption{$t$-channel}
\label{fig:t-channel s s O}
\end{subfigure}
~
\begin{subfigure}[t]{.4\textwidth}
\centering
\begin{tikzpicture}[xscale=0.6,yscale=0.6]
\draw[thick] (2,0) arc (270:90:2);
\draw[thick] (2,0) -- (4,0);
\draw[thick] (2,4) -- (4,4);
\draw[thick, dashed] (0,2) -- (1.5,2);
\node[right] at (4,0) {$\s$};
\node[right] at (4,4) {$\s$};
\node[above] at (.75,2) {$\cO$};
\filldraw[gray] (2,2) circle (.5);
\end{tikzpicture}
\caption{$s$-channel}
\label{fig:s-channel s s O}
\end{subfigure}
\caption{An illustration of how the inversion formula relates between $s$- and $t$-channels in the $\langle \s\s\rangle_\beta$ correlator. A single term in the $t$-channel OPE $\cO\in \s\times \s$ represented in (a), inverts to a part of the sum over
the $[\s\s]_n$ families 
in the $s$-channel, which are represented in (b). Alternatively, the sum in  (b) over $a_{[\s\s]_n}^{\<\s\s\>, (\cO)}$ reproduces the $\cO$ term in (a).}
\label{fig:s s O crossing}
\end{figure}

We begin by solving for the lowest-twist family of operators in the theory, $[\s\s]_0$. The most direct way to study this family is through the $\langle \s\s\rangle_\beta$ two-point function. Large-spin perturbation theory instructs us to start by inverting the lowest-twist operators in the $t$-channel. The first few low-twist primary operators in the $\s\times\s$ OPE are
\begin{align}
\s\times \s = \mathbf{1}  + T +\sum_{\ell=4,6,\dots} [\s\s]_{0,\ell}+ \epsilon+\dots.
\end{align} 
Note that $[\s\s]_0$ operators are nearly killed by $\mathrm{Disc}$ and thus give smaller contributions than $\mathbf{1},T,\e$.  Thus, we will initially neglect them, but we will add them in later. We have singled out $T$ from the rest of the $[\s\s]_0$ family because it has the largest anomalous dimension of the family and gives the least suppressed contribution.
Inverting the operators $\mathbf{1}$, $\e$, and $T$, we obtain a first approximation for $a_{[\s\s]_0}^{\langle\s\s\rangle}(J)$ \cite{Iliesiu:2018fao}
\begin{align}
a_{[\s\s]_0}^{\langle\s\s\rangle}(J) \supset \sum_{\cO = \mathbf{1},\e,T} a_\cO^{\langle\s\s\rangle} (1+(-1)^J) \frac{ K_J}{K_{\ell_\cO}} \pdr{\bar h}{J}S_{h_\cO-\Delta_\s,\Delta_\s}(\bar h). \label{eq:[ss]_0 leading one point}
\end{align} 
These contributions can be represented by the large-spin diagrams in figure~\ref{fig:s s O crossing}.

The next most significant contribution comes from the $[\s\s]_0$ family itself. To compute their contributions, one needs to sum over the family in the $t$-channel before inverting, as discussed in \cite{Iliesiu:2018fao}. The sum we need to do is\footnote{Note that terms with $s>\ell$ are absent from the $t$-channel sum, so for sufficiently large $s$, we need to start the sum at higher $\ell$. We ensure this by letting the sum start at $\ell= \min(\ell_0,s)$.}
\be
\sum_{s=0}^{\infty} \sum_{\ell=\min(\ell_0,s)}^\infty p_s(\ell) a_{[\s\s]_0}^{\langle\s\s\rangle}(\hb) (1-z)^{h(\hb)-2h_\s+s} (1-\zb)^{\hb-2h_\s-s}\, ,
\ee where $h(\hb) = 2h_\s + \delta_{[\s\s]_0}(\hb)$ and $\hb=h(\hb)+\ell$.
The sum is evaluated by expanding in small $\delta(\hb) \log (1-z)$, and then regulating the asymptotic parts of the $\hb$ sum, as was explained in \cite{Iliesiu:2018fao}. For the convenience of the interested reader, we review the method in appendix~\ref{appendix: alpha sums}. The result is as follows, 
\begin{align}
&\sum_{\ell=\ell_0}^\infty p_s(\ell)\, a_{[\s\s]_0}^{\langle\s\s\rangle}(\hb) (1-z)^{2h_\s+\delta_{[\s\s]_0}(\hb)-2h_\s+s} (1-\zb)^{\hb-2h_\s-s} \nn
\\ & = \sum_{m=0}^\infty \left( \sum_{a\in A_m} c_{a}\!\left[\frac{\delta_{[\s\s]_0}^m}{m!} p_s a_{[\s\s]_0}^{\langle\s\s\rangle} \right] \zb^a + \sum_{k=0}^\infty \alpha_k\!\left[\frac{\delta_{[\s\s]_0}^m}{m!} p_s a_{[\s\s]_0}^{\langle\s\s\rangle},\delta_{[\s\s]_0},2h_\s+s\right]\!\!(\hb_0) \, \zb^k\right) \nn
\\ & \qquad \x (1-z)^s \log^m (1-z),
\label{eq:t-chan ss0 sum}
\end{align} where $\hb_0=2h_\s+\ell_0$. Here, the set $A_m \subset \R \backslash \Z_{\ge 0}$ and the coefficients $c_a[f]$ are determined by the large-$\hb$ expansion of the summand $f(\hb)$, via \eqref{eq:S asymptotics of summand}.\footnote{We note that the terms $\zb^a$ can also include terms of the form $\zb^a \log^m \zb$ for $m\in \Z_{\ge0}$.} The coefficients $c_a[f]$ do not depend on the finite part of the sum. The coefficients $\alpha_k$ are computed via the formula \eqref{eq:def:alpha_k}, and depend on the details of the sum. We call the terms $\zb^a$ (and $\zb^a \log^m \zb$) `singular' terms, and the $\zb^k$ `regular' terms. The singular terms have are characterized by having nonzero $s$-channel discontinuity (near $\zb\sim 0$), while the regular terms have vanishing discontinuity.

The self-corrections of the $[\s\s]_0$ family are determined by the $k=0$ term on the right hand side. We are only interested in the leading large-$\hb$ contribution; recalling that the power of $\hb$ is controlled by the power of $(1-z)$, we need only consider the term with $s=0$. 
Taking the leading thermal coefficients in \eqref{eq:[ss]_0 leading one point} and summing over the $[\s\s]_0$ family starting at spin 4, and inverting, we obtain the first iteration of their self-correction;
\begin{align}
a_{[\s\s]_0}^{\langle\s\s\rangle}(J ) \supset &\sum_{\cO = \mathbf{1},\e,T} a_\cO^{\langle\s\s\rangle} (1+(-1)^J) \frac{K_J}{K_{\ell_\cO}} \frac{d\bar h}{dJ} \nonumber
\\ & \times \left(S_{h_\cO-\Delta_\s,\Delta_\s}(\bar h) + \sum_{m=0}^{\infty}\alpha_0^{\text{even}}\! \left[\frac{\delta^m_{[\s\s]_0}}{m!} S_{h_\cO-\Delta_\s,\Delta_\s},\delta_{[\s\s]_0}, \Delta_\s \right]\!(2h_\s+4) \, S^{(m)}_{0,\Delta_\s}(\bar h) \right). \label{eq:[ss]_0 self corrected}
\end{align} 
Note that $S_{0,\Delta}^{(0)}(\hb)=0$, so self-corrections start at order $\delta_{[\s\s]_0}$ and are suppressed by powers of the small anomalous dimensions. To evaluate the $\alpha$-sum above, we need the large-spin expansion of the $[\s\s]_0$ anomalous dimensions reproduced in \eqref{eq:delta ss0}. 
Concretely, the first few terms in the large-$\hb$ expansion are
\begin{align}
\delta_{[\s\s]_0}(\bar h)
\sim -0.001423 \frac{1}{\bar h}-0.04628 \frac{1}{\bar h^{\Delta_\e}} +\dots\,.
\label{eq:delta ss0 leading}
\end{align}

In principle, we can iterate the self-correction indefinitely. The solution to this iteration is the fixed-point of the self-correction map. How to solve for this fixed-point was also explained in \cite{Iliesiu:2018fao}. In practice, one needs to truncate to some order in the anomalous dimension expansion. Truncating to order $\delta^2$, the self-corrected thermal coefficients are
\begin{align}
a_{[\s\s]_0}^{\langle\s\s\rangle}(J ) &=  (1+(-1)^J) 4 \pi K_J \frac{d\bar h}{dJ} \nonumber
\\ & \quad\times \bigg(a_{\mathbf{1}}^{\langle\s\s\rangle}  \left( S_{-\Delta_\s,\Delta_\s}(\bar h) - 0.0119 \, S^{(1)}_{0,\Delta_\s}(\bar h) + 2.14 \times 10^{-5} S^{(2)}_{0,\Delta_\s}(\bar h) \right)  \nonumber
\\ & \qquad+ a_{\epsilon}^{\langle\s\s\rangle}  \left( S_{h_\e-\Delta_\s,\Delta_\s}(\bar h) + 0.0007999 \, S^{(1)}_{0,\Delta_\s}(\bar h)-1.95 \times 10^{-6} S^{(2)}_{0,\Delta_\s}(\bar h) \right) \nonumber
\\ & \qquad + a_{T}^{\langle\s\s\rangle} \frac{3}{8}  \left( S_{h_T-\Delta_\s,\Delta_\s}(\bar h) - 0.0001312 \, S^{(1)}_{0,\Delta_\s}(\bar h)+3.01 \times 10^{-7} S^{(2)}_{0,\Delta_\s}(\bar h) \right)  \bigg) \nn
\\ &\quad +\dots\,, \label{eq:[ss]_0 self corrected fixed point}
\end{align} where the dots denote terms suppressed in large-$\hb$ or in small $\delta_{[\s\s]_0}$. For convenience, plots of the three terms are given by the dashed curves in figure \ref{fig:ss0 thermal coefficents}.

\subsection{$[\s\s]_1$ and $[\e\e]_0$}

The next families that we solve for require more care. First, we compute the leading contributions to their thermal coefficients in the large-spin limit. Afterwards, we discuss subtleties that arise when considering finite spin members of the two families.

\subsubsection{Tree level contributions}

We start by computing the asymptotic contributions. Inverting the low-twist operators $\mathbf{1}$, $\e$, $T$, and the $[\s\s]_0$ family in the $\langle \s\s \rangle$ correlator gives `tree-level' contributions to the thermal coefficients of the $[\s\s]_1$ family. We can compute the contributions of $\mathbf{1}$, $\e$, and $T$ via \eqref{eq:residue of single t-chan op inversion} as
\begin{align}
&a_{[\s\s]_{1}}^{\langle\s\s\rangle,(\cO)} ( J) %= -\Res\limits_{\Delta = 2\Delta_s +2n+J} a^{(\cal O)}(\Delta,J)
\nn
\\ &\quad = a_\cO^{\langle\s\s\rangle} (1+(-1)^J) 4\pi K_J \frac{d\bar h}{dJ} \sum_{r=0}^{1} \sum_{s=0}^{\ell_{\cal O}} q_r(J) p_s(\ell_\cO) (-1)^{n-r} \binom{\bar h_\cO-\Delta_\s -s}{n-r} 
S_{h_\cO-\Delta_\s+s,\Delta_\s-r}(\bar h) \nn
\\ &\quad =a_\cO^{\langle\s\s\rangle} (1+(-1)^J) \frac{K_J}{K_{\ell_\cO}} \frac{d\bar h}{dJ} \left(-(\hb_\cO -\Delta_\s) S_{h_\cO-\Delta_\s,\Delta_\s}(\hb) -\frac{2+J}{3+2J} S_{h_\cO-\Delta_\s,\Delta_\s-1}(\hb)\right) +\dots \label{eq:aO for ss1}
\end{align}  where the dots denote higher order terms in $1/\hb$ that we will drop. We can also sum over the rest of the $[\s\s]_0$ family and compute it's contribution to the $[\s\s]_1$ pole, similarly to how we computed the $[\s\s]_0$ self-correction in \eqref{eq:[ss]_0 self corrected}. Their leading contribution is given by 
\begin{align}
& a_{[\s\s]_{1}}^{\langle\s\s\rangle,([\s\s]_0)}(J) \nn
\\ &\quad = (1+(-1)^J) 4\pi K_J \frac{d\bar h}{dJ}\sum_{r=0}^{1} \sum_{m=0}^{\infty} q_r(J) \alpha_{n-r}%^{\text{even}}
\! \left[\frac{\delta^m_{[\s\s]_0}}{m!} p_0 a_{[\s\s]_0}^{\langle\s\s\rangle},\delta_{[\s\s]_0}, \Delta_\s \right]\!(2h_\s+4) \,  S_{0,\Delta_\s-r}^{(m)}(\hb). \label{eq:ss0 contribution to ss1}
%\\ &= \sum_{\cO=\mathbf{1},\e,T} a_\cO (1+(-1)^J) \frac{K_J}{K_{\ell_\cO}} \frac{d\bar h}{dJ} \sum_{r=0}^{1} \sum_{m=0}^{\infty} q_r(J) \alpha_{n-r}^{\text{even}}\! \left[\frac{\delta^m_{[\s\s]_0}}{m!} S_{h_\cO-\Delta_\s,\Delta_\s},\delta_{[\s\s]_0}, \Delta_\s \right]\!(2h_\s+4) \,  S_{0,\Delta_\s-r}^{(m)}(\hb)
\end{align}
Thus, by adding terms from \eqref{eq:aO for ss1} with those from \eqref{eq:ss0 contribution to ss1}, we find that at large spin
\begin{align}
a_{[\s\s]_{1}}^{\langle\s\s\rangle}(J) = \sum_{\cO = \mathbf{1},\e,T,[\s\s]_0} a_{[\s\s]_{1}}^{\langle\s\s\rangle,(\cO)}(J) +\dots.
\label{eq:ss1 at large spin}
\end{align}

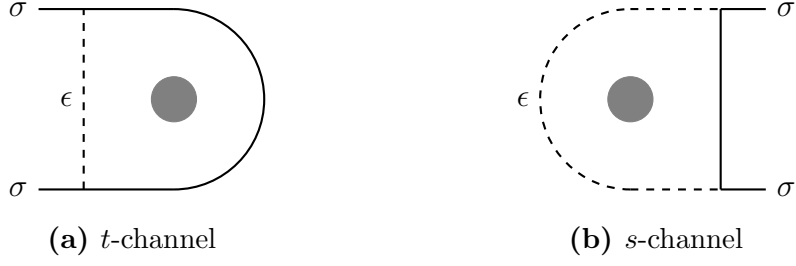
\begin{figure}[tb]
\centering
\begin{subfigure}[t]{.4\textwidth}
\centering
\begin{tikzpicture}[xscale=0.6,yscale=0.6]
\draw[thick] (0,0) -- (3,0);
\draw[thick] (0,4) -- (3,4);
\draw[thick, dashed] (1,0) -- (1,4);
\node[left] at (0,0) {$\s$};
\node[left] at (0,4) {$\s$};
\node[left] at (1,2) {$\e$};
\filldraw[gray] (3,2) circle (.5);
\draw[thick] (3,0) arc (-90:90:2);
\end{tikzpicture}
\caption{$t$-channel}
\label{fig:t-channel s s with delta}
\end{subfigure}
~
\begin{subfigure}[t]{.4\textwidth}
\centering
\begin{tikzpicture}[xscale=0.6,yscale=0.6]
\draw[thick,dashed] (2,0) arc (270:90:2);
\draw[thick,dashed] (2,0) -- (4,0);
\draw[thick,dashed] (2,4) -- (4,4);
\draw[thick] (4,0) -- (5,0);
\draw[thick] (4,4) -- (5,4);
\draw[thick] (4,0) -- (4,4);
\node[right] at (5,0) {$\s$};
\node[right] at (5,4) {$\s$};
\node[left] at (0,2) {$\e$};
\filldraw[gray] (2,2) circle (.5);
\end{tikzpicture}
\caption{$s$-channel}
\label{fig:s-channel s s with delta}
\end{subfigure}
\caption{The asymptotic parts of the $t$-channel sum over $[\s\s]_n$ represented by the diagram on the left inverts to the $s$-channel process on the right. Accordingly, their inversion should produce poles for the $[\e\e]_m$ families. The diagram on the left is deciphered by reading it from left to right; first the external $\s$ operators fuse into $[\s\s]_n$ states, which exchange an $\epsilon$ to correct their self-energy (anomalous dimension), then they receive expectation values proportional to $ b_{\mathbf{1}}$. The diagram on the right can also be deciphered by reading it from right to left; first the external $\s$ operators form $[\e\e]_m$ via exchange of a $\s$, then the $[\e\e]_m$ receive expectation values proportional to $ b_{\mathbf{1}}$.}
\label{fig:s s producing [ee]}
\end{figure}

What about the $[\e\e]_0$ family? The sum over the $[\s\s]_0$ family inside the $\langle \s\s\rangle$ correlator also contribute to the $[\e\e]_0$ family. Concretely, the sum over the $[\s\s]_0$ family contains asymptotics that sum to a `singular term' that corresponds to a pole for the $[\e\e]_0$ family. We can see this by the large spin diagrams in figure~\ref{fig:s s producing [ee]}. This gives a contribution
\begin{align}
a_{[\e\e]_0}^{\langle\s\s\rangle}(J) \supset (1+(-1)^J) 4 \pi K_J \, a_{\mathbf{1}}^{\langle\s\s\rangle}\, \delta_{[\s\s]_0}^{(\e)} \frac{\Gamma(\De_\s-\De_\e)}{\Gamma(\De_\s)} \, S_{0,\Delta_\s}^{(1)}(\hb). \label{eq:ee0 contribution from ss0 delta}
\end{align} Here, we have used the coefficient $\delta_{[\s\s]_0}^{(\e)}$ in the large-$\hb$ expansion of the anomalous dimension,
\begin{align}
\delta_{[\s\s]_0}(\hb) = \sum_{\cO} \delta_{[\s\s]_0}^{(\cO)} \frac{1}{\hb^{2h_\cO}}\, ,
\end{align} with the first few coefficients given in \eqref{eq:delta ss0} and \eqref{eq:delta ss0 leading}. Of course, this is only a naive approximation of  the $[\e\e]_0$ thermal coefficients which should only work for very large $J$. The $[\e\e]_0$ family is more directly accessed in the $\langle \e\e \rangle$ correlator, where inverting any single operator gives direct contribution to this family. For example, inverting the low-twist operators $\mathbf{1}$, $\e$, and $T$ in the $\langle \e\e \rangle$ correlator gives
\begin{align}
a_{[\e\e]_0}^{\langle \e\e\rangle}(J) \supset \sum_{\cO = \mathbf{1},\e,T} a_\cO^{\langle\e\e\rangle} (1+(-1)^J) \frac{ K_J}{K_{\ell_\cO}} \pdr{\bar h}{J}S_{h_\cO-\Delta_\e,\Delta_\e}(\bar h). \label{eq:[ee]_0 leading one point}
\end{align} Here, we labeled the thermal coefficients to indicate that they are the coefficients in the $\langle \e\e\rangle$ correlator. The relation between the thermal coefficients in the two correlators is given by the ratio of the OPE coefficients,
\begin{align}
a_{\cO}^{\langle\s\s\rangle} = \frac{f_{\s\s\cO}}{f_{\e\e\cO}} a_{\cO}^{\langle\e\e\rangle}.
\end{align}

Combining our result \eqref{eq:[ss]_0 self corrected fixed point} for $a_{[\s\s]_0}^{\langle\s\s\rangle}$ from $\langle \s\s \rangle$ with the ratio of OPE coefficients $f_{\s\s[\s\s]_0}/f_{\e\e[\s\s]_0}$ obtained from the analytic four-point function bootstrap, we can consider the contributions of the $[\s\s]_0$ family in the $\langle \e\e\rangle$ correlator. For example, their contribution to the $[\e\e]_0$ thermal coefficients can be computed, correcting \eqref{eq:[ee]_0 leading one point} as
\begin{align}
a_{[\e\e]_0}^{\langle \e\e\rangle}(J) \supset &\sum_{\cO = \mathbf{1},\e,T} a_\cO^{\langle\e\e\rangle} (1+(-1)^J) \frac{K_J}{K_{\ell_\cO}} \frac{d\bar h}{dJ} S_{h_\cO-\Delta_\e,\Delta_\e}(\bar h) \nonumber
\\ & + (1+(-1)^J) K_J \frac{d\bar h}{dJ}\sum_{m=0}^{\infty}\alpha_0\! \left[\frac{\delta^m_{[\s\s]_0}}{m!} \frac{f_{\e\e[\s\s]_0}}{f_{\s\s[\s\s]_0}} p_0 a_{[\s\s]_0}^{\langle\s\s\rangle},\delta_{[\s\s]_0}, \Delta_\s \right]\!(2h_\s+4) \, S^{(m)}_{\De_\s-\De_\e,\Delta_\s}(\bar h) . \label{eq:naive a_[ee]_0}
\end{align}

While at large $\hb$, \eqref{eq:ss1 at large spin} and \eqref{eq:naive a_[ee]_0} provide good approximations for the thermal coefficients this will not be the case at small $\hb$. In this regime, the two families $[\s\s]_1$ and $[\e\e]_0$ are very close together in twist, and have very large anomalous dimensions due to the operator mixing described in section \ref{sec:3dIsingReview}. Na\"ively, since the families are so close in twist, and strongly mix, we simply can't be sure how the residues are distributed between the families. More systematically, the presence of large anomalous dimensions means that the poles for the families are actually quite far from the na\"ive locations at $h=2h_\s+1$ and $2h_\e$ that were used to obtain \eqref{eq:ss1 at large spin} and \eqref{eq:naive a_[ee]_0}.  The effects that produce anomalous dimensions also produce corrections to the residues on a similar scale; since the anomalous dimensions are large at these intermediate $\hb$ values, the contributions to the residue must also be similarly large. Finally, there are altogether other poles for multi-twist families near the twists of these families, which the residues could further mix with.

We need to develop an approach to estimate the correct, mixed thermal coefficients. In order to estimate the correct, mixed thermal coefficients, we thus need to take into account all the corrections mentioned above. Towards that end, we now turn to developing some required technology.

\subsection{The half-inverted correlator}
\label{sec:half-invert}

Each individual $t$-channel block contributes only double-twist poles in the $s$-channel. However, the physical correlator has poles at non-double-twist locations. Consequently, the sum over $t$-channel blocks cannot commute with the inversion integral when $\De$ is near the physical poles. To see why, consider a contour integral around the location of a physical pole in $\De$. This integral gives zero for every $t$-channel block, but is certainly nonzero for the full $a(\De,J)$. By contrast, the sum over $t$-channel blocks does commute with the inversion integral when $\De$ is imaginary. However, we would like to determine numerically what happens at real $\De$.

To get a better numerical handle on how poles can shift, we will work with a more convenient object than $a(\Delta,J)$. Let's imagine applying the inversion formula `halfway', where we do the $z$ integral to compute the residues, but leave the $\zb$ integral --- which produces the poles --- undone. We want to define a generating function of the form
\begin{align}
%\tilde a (\zb,\hb) %&= (1+(-1)^J) K_J \int_1^{2} \frac{dz}{z} (z \bar z)^{\De_\f-\frac \De 2-\nu}(z-\bar z)^{2\nu} F_J\p{\sqrt{\frac {\bar z} {z}}} \Disc[g(z,\bar z)]
%\\ 
%&= 
(1+(-1)^J) K_J \int_1^{2} \frac{dz}{z} \sum_{r=0}^\infty q_r(J) z^{\Delta_\phi-\bar h-r}\bar z^{\Delta_\phi+r} \Disc[g(z,\bar z)].  \label{eq:halfinvert naive}
\end{align} (Once again, we assume no contributions from the arcs of the inversion formula.)
Now, instead of poles in $h$, we have powers $\zb^h$. Furthermore, the anomalous dimension corrections to pole locations are of the form 
\begin{align}
\frac{\delta(\hb)^m}{m!} \zb^h \log^m \zb.
\end{align} 

The idea is that \eqref{eq:halfinvert naive}  
is almost the inverse Laplace transform in $h$ of 
\begin{align} a(h,\hb) = a(\Delta=\hb+h,J=\hb-h)\end{align}
 --- almost due to the pesky factor of $K_J$. The generating function we want should relate to $a(h,\hb)$ along the lines of 
\begin{align}
\tilde a (\zb,\hb) = - \oint \frac{dh}{2\pi i} \zb^h a(h,\hb) ,\label{eq:inverse Laplace transform}
\end{align} which is the inverse to
\begin{align}
a(h,\hb) = \int_0^1 \frac{d\zb}{\zb} \zb^{-h} \tilde a (\zb,\hb) .
\end{align}
The inverse Laplace transform (\ref{eq:inverse Laplace transform}) can be performed in a region of $h$ where the inversion integral commutes with the sum over $t$-channel blocks, and thus we expect it to have a convergent expansion in $t$-channel blocks. The idea of defining a ``half-inverted" correlator was discussed in the four-point function case in \cite{Simmons-Duffin:2016wlq,Caron-Huot:2017vep}.

 The definitions \eqref{eq:halfinvert naive} and \eqref{eq:inverse Laplace transform} will agree if we make a few small modifications. Firstly, we should of absorb the factor of $K_J$ inside $a(h,\hb)$, so the contour integral in \eqref{eq:inverse Laplace transform} does not pick up unwanted poles. 
(At small enough twist $h$ such that we are away from poles in $K_J$, we can skip this step.) 
Secondly, we should reinterpret $J$ in $\tilde a (\zb,\hb)$ as an appropriate differential operator, $\hat J$, as we will explain below. Thus, we define
\begin{align}
\tilde a (\zb,\hb) %&= (1+(-1)^J) K_J \int_1^{2} \frac{dz}{z} (z \bar z)^{\De_\f-\frac \De 2-\nu}(z-\bar z)^{2\nu} F_J\p{\sqrt{\frac {\bar z} {z}}} \Disc[g(z,\bar z)]
%\\ 
&= \frac{1}{4\pi} \int_1^{2} \frac{dz}{z} \sum_{r=0}^\infty q_r(\hat J) z^{\Delta_\phi-\bar h-r}\bar z^{\Delta_\phi+r} \Disc[g(z,\bar z)],  \label{eq:halfinvert}
\end{align} which satisfies
\begin{align}
a(h,\hb) = (1+(-1)^J) 4\pi K_J \int_0^1 \frac{d\zb}{\zb} \zb^{-h} \tilde a (\zb,\hb) .
\label{eq:halfinvert to full invert}
\end{align} 
We call $\tilde a (\zb,\hb)$ the half-inverted correlator.

Inside half-inverted correlators, $J$ should be thought of as the linear operator 
\begin{align}
\hat J &= \hb - h =\hb - \zb \, \partial_{\zb}
\end{align} acting on the space of functions of the form $\zb^h \log^m \zb$. 
Note that $\hat J$ appears in $\tilde a(\zb,\hb) $ inside $q_r(\hat J)$, which are rational functions of $\hat J$ for each integer $r$. Therefore, we will need to invert $\hat J$ when acting on this space of functions. 
 For brevity, let's denote 
\begin{align}
|h,m\rangle \equiv \zb^h \log^m \zb.
\end{align}
For our purposes, $h>0$ and $m$ is a non-negative integer. For example, we have
\begin{align}
\zb \, \partial_{\zb} |h,m\rangle = h|h,m\rangle+ m|h,m-1\rangle.
\end{align} Then, expressions such as
\begin{align}
\frac{1}{c+ d\, \hat J} = \frac{1}{c+d(\hb - \zb \, \partial_{\zb})}
\end{align} can be interpreted as the inverse of the appropriate linear operator acting on this space of functions. Inverting the operator $\zb \, \partial_{\zb}$, we have
\begin{align}
(\zb\, \partial_{\zb})^{-1} | h, m\rangle = \frac{1}{h} \sum_{k=0}^m (-1)^k \frac{m!}{(m-k)!} \frac{1}{h^k} |h,m-k\rangle.
\end{align}
Similarly,
\begin{align}
\frac{1}{c+ d\, \hat  J} | h, m\rangle = \frac{1}{c+d\, (\hb -h)} \sum_{k=0}^m (-1)^k \frac{m!}{(m-k)!} \frac{1}{(c+d\, (\hb -h))^k} |h,m-k\rangle.
\end{align} With this interpretation, we can substitute $\hat J$ for $J$ as we did in \eqref{eq:halfinvert}, and define the half-inverted correlator as an honest function of $\zb$ and $\hb$ satisfying \eqref{eq:halfinvert to full invert}.

\subsubsection{Contributions to the half-inverted correlators $\<\widetilde{\s\s}\>$ and $\<\widetilde{\e\e}\>$}
\label{sec:contributions to halfinverted correlators}

Returning to the Ising model, by half-inverting our low-twist operators $\mathbf{1}$, $\e$, $T$, and the $[\s\s]_0$ family in the $\langle \s\s\rangle$ and $\langle \e\e\rangle$ correlators, we obtain leading-order-in-large-$\hb$ approximations to the respective half-inverted correlators $\langle \widetilde{ \s\s} \rangle(\zb,\hb)$ and $\langle \widetilde{\e\e}\rangle(\zb,\hb)$. The terms we compute include those that give the na\"ive $[\s\s]_1$ and $[\e\e]_0$ thermal coefficients \eqref{eq:ss1 at large spin} and \eqref{eq:naive a_[ee]_0}, but also include many other terms coming from the sum over the $[\s\s]_0$ family. 

We are not just limited to inverting the operators $\mathbf{1}$, $\e$, $T$, and the $[\s\s]_0$ family. While we do not know enough about any of the other families in the theory to compute all of their contributions, there are a special set of contributions that we can compute. In particular, while the regular terms $\alpha_k$ depend on such particulars of the family as anomalous dimensions and an exact sum over the thermal coefficients, the singular terms do not. The singular terms only depend on the asymptotic expansions. Furthermore, the leading contributions to the singular terms are to constant order in the anomalous dimensions, thus we can compute them without any knowledge of the anomalous dimensions. 
Therefore, we can essentially take a half-inverted correlator, and attempt to partially solve it in the large-$\hb$ regime. 
Let's say that the sum over $[\s\s]_0$ produced a term 
\be
p(\hb) \zb^{h_f} \subset \< \widetilde{\s\s}\>
\ee where $h_f$ is the asymptotic half-twist of a multitwist family $f$.
We can safely say that $p(\hb)$ is a part of the large-$\hb$ asymptotics of the thermal coefficient of the family $f$.
Now, the sum over the family $f$ in the $t$-channel includes a term
\be
\sum_{\cO\in f}  (1+(-1)^\ell) p(\hb) (1-\zb)^{\hb-2h_\s}(1-z)^{h_f+\delta_{f}(\hb)-2h_\s} 
&\supset \sum_{a\in A} c_a[p] \zb^a (1-z)^{h_f-2h_\s}\nn
\\ & \qquad + \cO(\delta_f) +\text{regular}. \label{eq:sum over multitwist}
\ee Note that we can determine the singular term $c_a[p(\hb)]$ without having to know about the small-$\hb$ behavior of the thermal coefficients of the family $f$, or the anomalous dimensions $\delta_f$! This is unlike the regular terms, which depend on knowing the small-$\hb$ behavior of the thermal coefficients as well as the anomalous dimensions. Inverting the singular term in \eqref{eq:sum over multitwist}, we obtain a contribution to the half inverted correlator
\be
\< \widetilde{\s\s}\> \supset c_a[p] S_{h_f-2h_\s,2h_\s}(\hb) \zb^{2h_\s+a}. \label{eq:augmentation by singular term from multitwist}
\ee
So, we take the half-inverted correlators $\<\widetilde{\s\s}\>$ and $\<\widetilde{\e\e}\>$ computed from the contributions of $\mathbf{1}$, $\e$, $T$, and $[\s\s]_0$, and augment them with the singular terms \eqref{eq:augmentation by singular term from multitwist} coming from all the asymptotics of thermal coefficients of other families that appear in them. 

In fact, these singular terms are crucial, and augmenting by them is a natural thing to do. For example, in order to reproduce known anomalous dimensions from the thermal inversion formula --- such as those of the $[\s\s]_0$ family --- one needs to sum over multi-twist families in the $t$-channel \cite{Iliesiu:2018fao}. 
The prototypical example of this process is illustrated in the thermal large-spin diagram in figure~\ref{fig:aOprime s s with delta}. Also, recovering the thermal coefficients of $[\s\s]_0$ in $\< \e\e\>$ requires summing over generically multitwist families that are generated in $\<\e\e\>$ by the sum over $[\s\s]_0$, as illustrated in figure~\ref{fig:ss0 in <ee>}. We will now briefly review these relevant processes. 

\subsubsection{Generating anomalous dimensions in $\<\s\s\>$}

\begin{figure}[tb]
\centering
\begin{subfigure}[t]{.4\textwidth}
\centering
\begin{tikzpicture}[xscale=0.6,yscale=0.6]
\draw[thick] (0,0) -- (3,0);
\draw[thick] (0,4) -- (3,4);
\draw[thick, dashed] (1,0) -- (1,4);
\node[left] at (0,0) {$\s$};
\node[left] at (0,4) {$\s$};
\node[left] at (1,2) {$\cO$};
\filldraw[gray] (3,2) circle (.5);
\draw[thick] (3,0) arc (-90:90:2);
\draw[thick, dotted] (3.5,2) -- (5,2);
\node[above] at (4.25,2) {$\cO'$};
\end{tikzpicture}
\caption{$t$-channel}
\label{fig:t-channel aOprime phi phi with delta}
\end{subfigure}
~
\begin{subfigure}[t]{.4\textwidth}
\centering
\begin{tikzpicture}[xscale=0.6,yscale=0.6]
\draw[thick,dashed] (2,0) arc (270:90:2);
\draw[thick,dashed] (2,0) -- (4,0);
\draw[thick,dashed] (2,4) -- (4,4);
\draw[thick] (4,0) -- (5,0);
\draw[thick] (4,4) -- (5,4);
\draw[thick] (4,0) -- (4,4);
\node[right] at (5,0) {$\s$};
\node[right] at (5,4) {$\s$};
\node[left] at (0,2) {$\cO$};
\filldraw[gray] (2,2) circle (.5);
\draw[thick, dotted] (2.5,2) -- (4,2);
\node[above] at (3.25,2) {$\cO'$};
\end{tikzpicture}
\caption{$s$-channel}
\label{fig:s-channel aOprime phi phi with delta}
\end{subfigure}
\caption{The $t$-channel diagram denotes a sum over the asymptotics $\delta_{[\s\s]_n}^{(\cO)}(\bar h) \times a_{[\s\s]_n}^{\langle\s\s\rangle(\cO')}(\bar h)$. This inverts to poles for the $[\cO\cO\cO']_m$ families in the $s$-channel. Conversely, swapping the $s$- and $t-$channels, and summing over the $[\cO\cO\cO']_m$ family in the $t$-channel reproduces the anomalous dimensions of the $[\s\s]_n$ family in the terms proportional to $a_{\cO'}$.} 
\label{fig:aOprime s s with delta}
\end{figure}
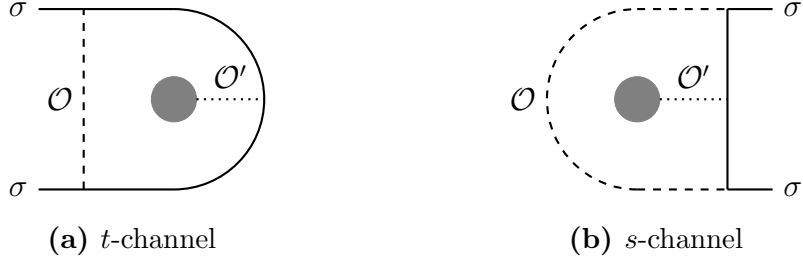

Let's illustrate how anomalous dimensions are generated for the half-inverted correlator $\<\widetilde{\s\s}\>$ by an example. We saw in \eqref{eq:ee0 contribution from ss0 delta} that the sum over $[\s\s]_0$ in $\langle\s\s\rangle$ produced a pole for the $[\e\e]_0$ family. In particular, this means that the sum over $[\s\s]_0$ contributes a term
\begin{align}
{\langle \widetilde{\s\s}\rangle}(\zb,\hb) &\supset %(1+(-1)^J) 4 \pi K_J 
\zb^{2h_\e} a_{\mathbf{1}}^{\langle\s\s\rangle}\, \delta_{[\s\s]_0}^{(\e)} \frac{\Gamma(\De_\s-\De_\e)}{\Gamma(\De_\s)} \, S_{0,\Delta_\s}^{(1)}(\hb) \label{eq:ee0 in ss half invert}
\end{align} to the half-inverted correlator ${\langle \widetilde{\s\s}\rangle}(\zb,\hb)$. This implies that there is a term, given in \eqref{eq:ee0 contribution from ss0 delta}, in the large-$\hb$ expansion of $a_{[\e\e]_0}^{\langle\s\s\rangle}(\hb)$. Now, we would be wrong to say that this is a good approximation to the thermal coefficients at small $\hb$, but at large $\hb$, we know such a term is there. By crossing symmetry of figure~\ref{fig:s s producing [ee]}, this term is responsible for generating the $\delta_{[\s\s]_0}^{(\e)}$ correction to the anomalous dimensions of $[\s\s]_0$ in $\langle \widetilde{\s\s}\rangle$. 

Let's consider the contributions of $[\e\e]_0$ to the thermal coefficients in $\langle \s\s\rangle$. To evaluate them, we need to analyze the $t$-channel sum over the family. This sum has the same form as the sum \eqref{eq:t-chan ss0 sum} over the $[\s\s]_0$ family, 
\begin{align}
\sum_{\ell=\ell_0}^\infty &p_0(\ell) a_{[\e\e]_0}^{\langle\s\s\rangle}(\hb) (1-z)^{2h_\e+\delta_{[\e\e]_0}(\hb)-2h_\s} (1-\zb)^{\hb-2h_\e} \nn
\\ & = \sum_{m=0}^\infty \left( \sum_{a\in A_m} c_a\!\left[\frac{\delta_{[\e\e]_0}^m}{m!} p_0 a_{[\e\e]_0}^{\langle\s\s\rangle} \right] \zb^a + \sum_{k=0}^\infty \alpha_k\!\left[\frac{\delta_{[\e\e]_0}^m}{m!} p_0 a_{[\e\e]_0}^{\langle\s\s\rangle},\delta_{[\e\e]_0},2h_\s\right](\hb_0) \zb^k\right) \nn
\\ &\qquad \x(1-z)^{2h_\e-2h_\s} \log^m (1-z),
\end{align} where $\hb = 2h_\e+\ell + \delta_{[\e\e]_0}(\hb)$ and $\hb_0=2h_\e + \ell_0$. One important difference is that since $2h_\e-2h_\s\notin \Z_{\ge 0}$, the terms with $m=0$ have nonzero discontinuity and contribute to the inversion formula. So, we can consider the leading term $m=0$ in the anomalous dimension expansion. 
Now, without knowledge of small-$\hb$ values of $a_{[\e\e]_0}^{\langle\s\s\rangle}(\hb)$, we can't reliably evaluate the $\alpha_k$ coefficients. However, the coefficients $c_a[p]$ only depend on the asymptotic expansion of $p(\hb)$, and are insensitive to small-$\hb$ behavior. So, using the term of $a_{[\e\e]_0}^{\langle\s\s\rangle}$ in \eqref{eq:ee0 in ss half invert}, we can compute the leading singular term $c_a\!\left[p_0 a_{[\e\e]_0}^{\langle\s\s\rangle}\right] \zb^a$,
\begin{align}
\sum_{a\in A_0} c_a\!\left[p_0 a_{[\e\e]_0}^{\langle\s\s\rangle}\right] \zb^a \supset a_{\mathbf{1}}^{\langle\s\s\rangle}\, \delta_{[\s\s]_0}^{(\e)} \frac{\Gamma(\De_\s-\De_\e)}{\Gamma(\De_\s)} \log \zb.
\label{eq:ee0 leading singular term in ss}
\end{align} Half-inverting this term, we obtain the corresponding contribution
\begin{align}
{\langle\widetilde{\s\s}\rangle}(\zb,\hb) \supset a_{\mathbf{1}}^{\langle\s\s\rangle}\, \delta_{[\s\s]_0}^{(\e)} \frac{\Gamma(\De_\s-\De_\e)}{\Gamma(\De_\s)} \zb^{2h_\s} \log \zb \, S_{2h_\e-2h_\s,2h_\s}(\hb) +\dots. \label{eq:ee0 augment to ss}
\end{align} This is a correction to the anomalous dimension (pole location) $\delta_{[\s\s]_0}$ of the $[\s\s]_0$ family. As expected, this is exactly the term in large-spin perturbation theory that produces the contribution of $\e$ to the anomalous dimension through the crossing-symmetric process illustrated in figure~\ref{fig:s s producing [ee]}. Other contributions arise from similar sums over other, potentially multi-twist families, as illustrated in figure~\ref{fig:aOprime s s with delta}.

One important point to highlight is that the contribution \eqref{eq:ee0 augment to ss} above does not only produce the expected anomalous dimension, it also contributes to higher poles. The half-inversion of the term in \eqref{eq:ee0 leading singular term in ss} produces another term, contributing to the anomalous dimensions at the na\"ive location of the $[\s\s]_1$ family,
\begin{align}
{\langle\widetilde{\s\s}\rangle}(\zb,\hb) \supset a_{\mathbf{1}}^{\langle\s\s\rangle}\, \delta_{[\s\s]_0}^{(\e)} \frac{\Gamma(\De_\s-\De_\e)}{\Gamma(\De_\s)} \zb^{2h_\s+1} \log \zb \, q_1(\hat J) S_{2h_\e-2h_\s,2h_\s-1}(\hb).
\end{align} In principle, this is an important contribution when considering the $[\s\s]_1$ family, and through mixing, the $[\e\e]_0$ family. The moral is that we should systematically generate these terms by iterating $t$-channel sums and subsequent half-inversions, rather than by putting the anomalous dimensions in by hand whenever they are known, as we will also generate other contributions. In summary, we put in the anomalous dimensions of $[\s\s]_0$ and recover them, but also generate some additional terms for $[\s\s]_n$.

\subsubsection{Generating $[\s\s]_0$ in $\<\widetilde{\e\e}\>$}
\label{sec:ss0 in <ee>}

Another important phenomenon is the generation of the $[\s\s]_0$ thermal coefficients in $\<\widetilde{\e\e}\>$. Using $\<\s\s\>$, we already computed an expression for the $[\s\s]_0$ thermal coefficients, which we believe to be accurate. One might be tempted to input them into $\<\widetilde{\e\e}\>$ by hand. As with the anomalous dimensions above, it's worthwhile to generate the $[\s\s]_0$ thermal coefficients in $\<\widetilde{\e\e}\>$ systematically; similarly, contributions to the $[\s\s]_1$ thermal coefficients in $\<\widetilde{\e\e}\>$ are also generated.

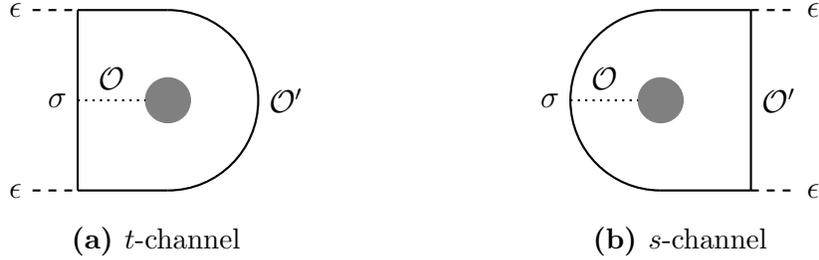
\begin{figure}[tb]
\centering
\begin{subfigure}[t]{.4\textwidth}
\centering
\begin{tikzpicture}[xscale=0.6,yscale=0.6]
\draw[thick,dashed] (0,0) -- (1,0);
\draw[thick,dashed] (0,4) -- (1,4);
\draw[thick] (1,0) -- (3,0);
\draw[thick] (1,4) -- (3,4);
\draw[thick] (1,0) -- (1,4);
\node[left] at (0,0) {$\e$};
\node[left] at (0,4) {$\e$};
\node[left] at (1,2) {$\s$};
\filldraw[gray] (3,2) circle (.5);
\draw[thick] (3,0) arc (-90:90:2);
\draw[thick, dotted] (1,2) -- (2.5,2);
\node[above] at (1.75,2) {$\cO$};
\node[right] at (5,2) {$\cO'$};
\end{tikzpicture}
\caption{$t$-channel}
\label{fig:t-channel ss0 in <ee>}
\end{subfigure}
~
\begin{subfigure}[t]{.4\textwidth}
\centering
\begin{tikzpicture}[xscale=0.6,yscale=0.6]
\draw[thick] (2,0) arc (270:90:2);
\draw[thick] (2,0) -- (4,0);
\draw[thick] (2,4) -- (4,4);
\draw[thick,dashed] (4,0) -- (5,0);
\draw[thick,dashed] (4,4) -- (5,4);
\draw[thick] (4,0) -- (4,4);
\node[right] at (4,2) {$\cO'$};
\node[right] at (5,0) {$\e$};
\node[right] at (5,4) {$\e$};
\node[left] at (0,2) {$\s$};
\filldraw[gray] (2,2) circle (.5);
\draw[thick, dotted] (0,2) -- (1.5,2);
\node[above] at (.75,2) {$\cO$};
\end{tikzpicture}
\caption{$s$-channel}
\label{fig:s-channel ss0 in <ee>}
\end{subfigure}
\caption{To obtain the $[\s\s]$ thermal coefficients proportional to $a_\cO^{\<\e\e\>}$ in the $s$-channel of $\<\e\e\>$, one must invert sums over $[\cO\cO'\cO']$ in the $t$-channel. 
Of course, the diagrams are crossing symmetric, so the required $t$-channel terms are obtained from inverting the sum over $[\s\s]$ in the first place.
}
\label{fig:ss0 in <ee>}
\end{figure}

The process with which the $[\s\s]_0$ thermal coefficients are generated in $\<\widetilde{\e\e}\>$ is depicted in figure~\ref{fig:ss0 in <ee>}. Our task boils down to looking at the singular terms arising from the sum over $[\s\s]_0$ in $\<\e\e\>$,
\be
&\sum_{\ell} p_0(\ell) \frac{f_{\e\e[\s\s]_0}(\hb)}{f_{\s\s[\s\s]_0}(\hb)} a_{[\s\s]_0}^{\<\s\s\>} (\hb) (1-z)^{2h_\s+\delta_{[\s\s]_0}(\hb)-2h_\e} (1-\zb)^{\hb-2h_\s} \nn
\\ &\qquad \supset (1-z)^{2h_\s-2h_\e} \sum_{m=0}^\infty \log^m(1-z)\sum_{a\in A_m} c_a\!\left[\frac{\delta_{[\s\s]_0}^m}{m!} p_0\frac{f_{\e\e[\s\s]_0}}{f_{\s\s[\s\s]_0}} a_{[\s\s]_0}^{\<\s\s\>}  \right] \zb^a,
\ee
and then considering the sum over the families appearing there. The singular terms of the sums over those families (to constant order in their anomalous dimensions) reproduce the $[\s\s]_0$ thermal coefficients we seek. As before, inverting anything that contributes to a pole for $[\s\s]_0$ at $h=2h_\s$ also contributes to higher poles at $h=2h_\s+n$, and in particular to $[\s\s]_1$.

\subsection{Mixing between families}
\label{section:mixing between families}

The combination of our effort so far allows us to compute good approximations for the half-inverted correlators ${\langle \widetilde{\s\s}\rangle}(\zb,\hb)$ and ${\langle \widetilde{\e\e} \rangle}(\zb,\hb)$. To summarize our steps so far, our approximations are obtained first by half-inverting $\mathbf{1}$, $\e$, $T$, and the $[\s\s]_0$ family, and then further refined by augmenting by the singular terms coming from sums over other families (that appear in ${\langle \widetilde{\s\s}\rangle}(\zb,\hb)$ and ${\langle \widetilde{\e\e} \rangle}(\zb,\hb)$ from the asymptotics of the sum over the $[\s\s]_0$ family). 
Let $\tilde g^c(\zb,\hb)$ denote the vector of half-inverted correlators 
\be
\tilde g(\zb,\hb) =\left( {\langle\widetilde{ \s\s}\rangle}(\zb,\hb), {\langle\widetilde{\e\e}\rangle}(\zb,\hb) \right), \,
  \ee
  where $c$ labels the correlator. Our computations for the half-inverted correlators produce approximations of the form
\begin{align}
\label{eq:barz-expansion}
\tilde g_\text{na\"ive}^c(\zb,\hb) = \sum_{f} (a^c_f)^\text{na\"ive}(\hb) \,\zb^{h_f^\text{na\"ive}} \left(1+ \delta_f(\hb) \log \zb + O(\log^2 \zb) \right)
\end{align}
for each of the two correlators, $c$. Here, the sum is over several of the low-twist families $f$, such as $[\sigma \sigma]_0$, $[\sigma \sigma]_1$, $[\epsilon \epsilon]_0$, and a few others appearing as singular terms from the sum over $[\s\s]_0$. At sufficiently high $\bar h$, the $\log \bar z$ terms, like those found in \eqref{eq:ee0 augment to ss}, correctly approximates the anomalous dimensions for some of these families.\footnote{Note that our approach does not lead to the expected $\log \bar z$ terms for every family $f$. This is one reason for which considering mixing proves important. } However, at small $\bar h$, the thermal coefficients of families that are close in twist --- and thus have similar powers of $\bar z$ in the expansion \eqref{eq:barz-expansion} --- prove difficult to disentangle. As reviewed in \ref{sec:3dIsingReview}, in the case of the 3D Ising CFT,  the contributions of $[\s\s]_1$ and $[\e\e]_0$ are difficult to disentangle as   $h^{\text{naive}}_{[\s\s]_1}= \Delta_\s + 1= 1.518$, while $h^{\text{naive}}_{[\e\e]_0}= \Delta_\e = 1.412$. 
For this reason, we cannot simply identify the one point functions and anomalous dimensions of each family from the expansion \eqref{eq:barz-expansion}. We will instead use the augmented half-inverted correlators from $\tilde g_{\text{naive}}$ to implement a mixing procedure that disentangles the contributions of the three most important double-twist families in the 3D Ising CFT: $[\s \s]_0$, $[\s\s]_1$, and $[\e \e]_0$.

Using the ingredients in section~\ref{sec:half-invert} we can now explain the mixing procedure. We expect a given half-inverted correlator to have the exact form \begin{align}
\tilde g^c(\zb,\hb) = \sum_{f} a^c_f(\hb) \zb^{h_f(\hb)}, %\left( 1+  \delta_f^c(\bar h) \log \bar z + O(\delta_f^c(\bar h)^2 \log^2 \bar z) \right) \,.
\label{eq:true half-inverted correlators}
\end{align}
where the sum is over families $f$ once again, with the thermal coefficients in each family given by $a^c_f(\hb)$ and the exact half-twist given by $h_f(\hb)$. In the 3d Ising CFT we would like to truncate the sum of families to $f \in \cF =  \{[\s\s]_0,\,[\s\s]_1,\,[\e\e]_0\}$, which, due to their low twist, have the greatest contribution to the two correlators $\langle\s\s\rangle$ and $\langle\e\e\rangle$ in the light-cone limit. We will denote these truncations $g^c_\cF(\zb,\hb)$. At small $\zb$, $g^c(\zb,\hb)$ is dominated by the families $f\in \cF$, and therefore well approximated by $g^c_\cF(\zb,\hb)$.

We do not include multi-twist families such as $[\s\s\e]$ and $[\s\s\s\s]$ in the sum over $f$ for two reasons. The first is that they give a small numerical contribution to the flat-space four-point functions $\<\s\s\s\s\>, \<\s\s\e\e\>, \<\e\e\e\e\>$, so it is reasonable to guess that their contribution to thermal two-point functions is also small. The other reason is that we know much less about their anomalous dimensions and OPE coefficients, and thus wouldn't be able to write a suitable ansatz anyway. It will be important to better understand multi-twist operators to improve our techniques in the future.

The thermal coefficients appearing in different correlators are related by ratios of OPE coefficients. For each family, let's pick a thermal coefficient $a^{u}(\hb)$ from a certain correlator that we'd like to parametrize the thermal data of that family by. Given our choice of $a^{u}(\hb)$, we can form the matrix $\lambda_{u}^{c}(\zb, \hb)$ comprised of appropriate ratios of OPE coefficients such that 
\begin{figure}
\centering{
\includegraphics[width=.7\textwidth]{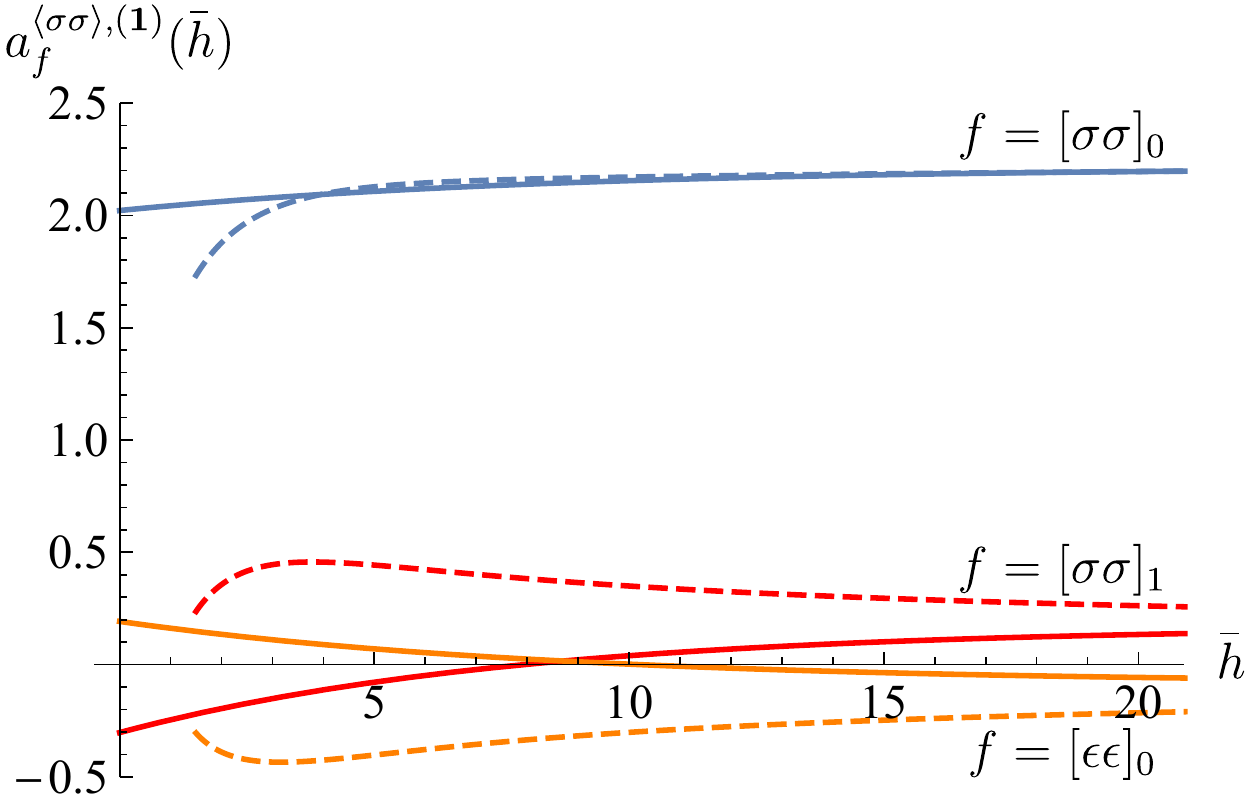}}
\caption{The effect of operator mixing for the thermal coefficients in the $[\s\s]_0$, $[\s\s]_1$, and $[\e\e]_0$ families. As an example we show the coefficient of $a_{\mathbf{1}}^{\<\s\s\>}$ in the thermal coefficients of each family. The dashed curves represent the predictions made by the inversion formula before implementing operator mixing, while the solid curves represent the post-mixing predictions, with the mixing region $\mathcal{P}_\text{mix}=\{0.05,0.1,\dots,0.3\}$.     }
\label{fig:mixing}
\end{figure}
\begin{align}
\label{eq:two-point-function-accurate}
\tilde g^c_\cF(\zb,\hb) = \lambda_{u}^{c}(\zb, \hb) a^{u}(\hb).
\end{align} 
Specifically, the exact contribution of the families $[\s\s]_0,\,[\s\s]_1,$ and $[\e\e]_0$ to the half-inverted correlator can be written using, 
\begin{align}
a^u (\hb) =\begin{pmatrix} a_{[\s\s]_0}^{\langle \s\s\rangle}(\hb) \\ a_{[\s\s]_1}^{\langle \s\s\rangle} (\hb) \\ a_{[\e\e]_0}^{\langle \e\e\rangle} (\hb) \end{pmatrix}.
\end{align}
Accordingly, we have
\begin{align}
\lambda^{c}_u = \begin{pmatrix}
\zb^{h_{[\s\s]_0}(\hb)} & \zb^{h_{[\s\s]_1}(\hb)} & \frac{f_{\s\s[\e\e]_0}(\hb)}{f_{\e\e[\e\e]_0}(\hb)} \, \zb^{h_{[\e\e]_0}(\hb)}
\\ \frac{f_{\e\e[\s\s]_0}(\hb)}{f_{\s\s[\s\s]_0}(\hb)} \, \zb^{h_{[\s\s]_0}(\hb)} & \frac{f_{\e\e[\s\s]_1}(\hb)}{f_{\s\s[\s\s]_1}(\hb)} \,\zb^{h_{[\s\s]_1}(\hb)} & \zb^{h_{[\e\e]_0}(\hb)}
\end{pmatrix}.
\end{align}
We can now understand Eq.~\eqref{eq:barz-expansion} as an approximation to the contribution of the families  correlator, 
\be 
\label{eq:mixing-ansatz}
\tilde g_\text{naive}^c(\zb,\hb) \approx \tilde g^c_\cF(\zb,\hb).
\ee 
Note that at large $\bar h$, due to the decrease in the anomalous dimensions for all three families in $\mathcal F$, the terms $(a^c_f)^{\text{naive}}(\bar h)$ appearing in \eqref{eq:barz-expansion} are close to the correct thermal coefficients  appearing in \eqref{eq:two-point-function-accurate}. However, at small values of $\bar h$, as has been described in section~\ref{sec:3dIsingReview}, the anomalous dimensions of operators in the $[\s\s]_1$ and $[\e\e]_0$ become large and thus there is a large $\bar z$-power mismatch between the terms which $(a^c_f)^{\text{naive}}(\bar h)$ in \eqref{eq:barz-expansion} and those that include $a^c_f(\hb) $ in \eqref{eq:true half-inverted correlators}. Thus, all the terms in the naive expansion  \eqref{eq:barz-expansion} will mix and contribute to the accurate thermal coefficients for all three families in $\mathcal F$. As previously mentioned, this effect is especially noticeable on families such as $[\s\s]_1$ and $[\e\e]_0$ whose twists are close and whose naive contribution in \eqref{eq:barz-expansion} are difficult to distinguish at small $\bar h$. For this reason, we will refer to \eqref{eq:mixing-ansatz} as the mixing equation.

\begin{figure}
\centering{
\includegraphics[width=.7\textwidth]{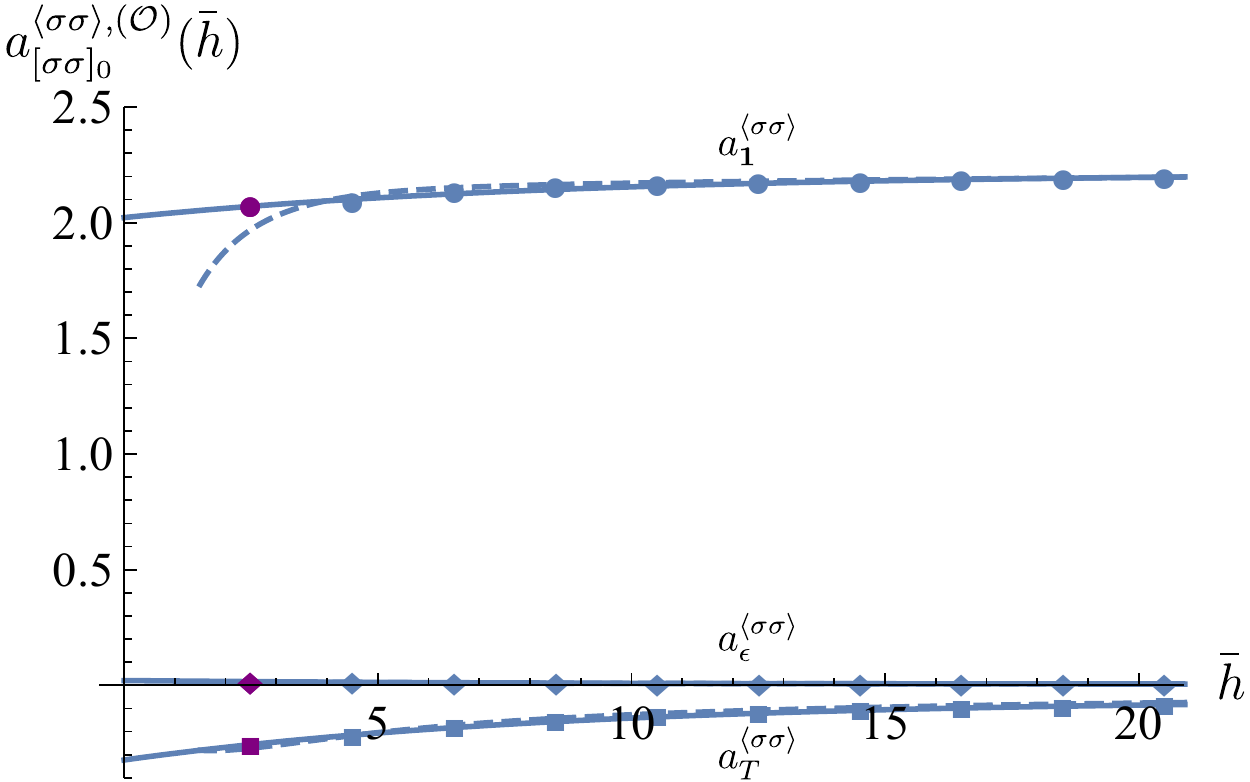}}
\caption{Estimates for the terms multiplying $a_\mathbf{1}^{\<\s\s\>}$, $a_\e^{\<\s\s\>}$, and $a_T^{\<\s\s\>}$ in the thermal coefficients $ a_{[\s\s]_0}^{\<\s\s\>}(J)$. The dashed blue curves are the predictions from the inversion formula before performing operator mixing while the solid curves are the predictions after accounting for operator mixing. The blue dots represent the post-mixing predictions for each local operator with $J\ge4$ in the $[\s\s]_0$ family. The purple dots are the extrapolation of the thermal coefficient to the stress-energy tensor.   }
\label{fig:ss0 thermal coefficents}
\end{figure}

In solving for the mixed coefficients $a^u(\bar h)$ we have conveniently written \eqref{eq:mixing-ansatz} in matrix form. Thus, for each value of $\bar h$ that we are interested in, we can treat the mixing equation as an over-determined linear system. Concretely, we can impose that \eqref{eq:mixing-ansatz} be satisfied for several values of $\bar z $ from some set of values $\mathcal P_{\text{mix}}$. Of course, due to the truncation of the expansion \eqref{eq:barz-expansion}, we get an overdetermined system of equations and it is impossible to satisfy the mixing equation for all values of $\bar z$. However, as one can see from figure \ref{fig:thermal coefficients of families}, when choosing, 
\be
\label{eq:mixing-zbar-vals}
\mathcal P_{\text{mix}} =  \{0.05,0.1,\dots, \bar z_\text{max} \}, \qquad \text{with}\quad  \bar z_\text{max}\in \{0.15,0.2,\dots, 0.6\}.
\ee
 our results are robust under different choices of $\mathcal P_{\text{mix}}$ (see figure \ref{fig:thermal coefficients of families}).\footnote{This remains true as long as $\bar z \gg O(1) e^{-1/\delta_{\cO}}$, where $\delta_\cO$ is the average anomalous dimension at a certain value of $\bar h$ for the three operator families that we are considering.} 
Thus, we solve for each term proportional to each unknown $a_{\mathbf{1},\, \epsilon,\,T}^{c}$ in $a^u (\hb)$  using the method of least squares for each value of $\bar h$.\footnote{We give an equal weight to each value of $\bar z$ in the least square fit.} To exemplify our procedure, in figure \ref{fig:mixing}, we show how the coefficients multiplying $a_{\mathbf{1}}^{\<\s\s\>}$ are affected by mixing.

We now use the estimates obtained from mixing to understand the thermal coefficients of operators with small spin. Since $T$ is a member of the $[\s\s]_0$ family, we can use our calculation of $a_{[\s\s]_0}^{\langle\s\s\rangle}$ to constrain $a_T^{\langle\s\s\rangle}$. We thus extrapolate our results for the thermal coefficients of  the $[\s\s]_0$ family down to $J=2$ (see figure \ref{fig:ss0 thermal coefficents}). After mixing, the thermal coefficient of $T$ is computed in terms of the unknowns as
\begin{align}
a_{[\s\s]_0}^{\langle\s\s\rangle}(\hb=2.5) = \left(\frac{d\hb}{dJ} \right) \bigg|_{\bar h=2.5}\left( 2.07 a_{\mathbf{1}}^{\langle\s\s\rangle} +0.0163 a_T^{\langle\s\s\rangle}-0.257 a_\epsilon^{\langle\s\s\rangle}\right)\,. \label{eq:aT from ss0 family curve}
\end{align}
 Using the known anomalous dimensions for the $[\s\s]_0$ family, we can compute $d\bar h/dJ$.\footnote{Since \cite{Simmons-Duffin:2016wlq} provides accurate values for the anomalous dimensions of all operators in $[\s\s]_0$, $[\s\s]_1$, and $[\e\e]_0$, we can use a fit to the numerical results to  accurately obtain  $d\bar h/dJ$. At the $\hb$ values of local operators, the fit strongly agrees with the analytical predictions for the anomalous dimensions. } 
Of course, \eqref{eq:aT from ss0 family curve} should be equal to $a_T^{\langle\s\s\rangle}$ itself! Solving for $a_T^{\langle\s\s\rangle}$, we have
\begin{align}
a_T^{\langle\s\s\rangle} = 2.136 a_{\mathbf{1}}^{\langle\s\s\rangle} - 0.265 a_\e^{\langle\s\s\rangle}.
\end{align} Recall that we can normalize all the thermal coefficients by that of the unit operator, thus setting $a_{\mathbf{1}}^{\langle\s\s\rangle}=1$. Therefore, we have only a single unknown left: $a_\e^{\langle\s\s\rangle}$. We have successfully approximated the thermal coefficients of all operators in the three low-twist families of interest in terms of a single unknown!

A similar issue presents itself when one considers low-spin operators in the higher-twist families $[\s\s]_1$ and $[\e\e]_0$. At spin 0 and 2, there are only the two operators $\e'$ and $T'$; both belong to the $[\s\s]_1$ family, whereas the $[\e\e]_0$ family has no such operators
\cite{Simmons-Duffin:2016wlq}. Therefore, our mixing procedure does not work for these operators. However, it's crucial to estimate the thermal coefficients of $\e'$ and $T'$ for solving the KMS condition. We have found it best to extract the thermal coefficients of the low-spin members of the $[\s\s]_1$ family by extrapolating the mixed thermal coefficients down to small $\hb$ by a simple fit. This is motivated by results from the flat-space data where the OPE coefficients and anomalous dimensions of these two operators appear to lie on smooth curves with all other members of the $[\s\s]_1$ family. The estimates for $a_{\e'}^{\<\s\s\>}$ and $a_{T'}^{\<\s\s\>}$ obtained by performing such a fit can be extrapolated using figure \ref{fig:mixing}.

\subsection{Solving for $b_\cO$}

\begin{figure}
\centering{
\includegraphics[width=.55\textwidth]{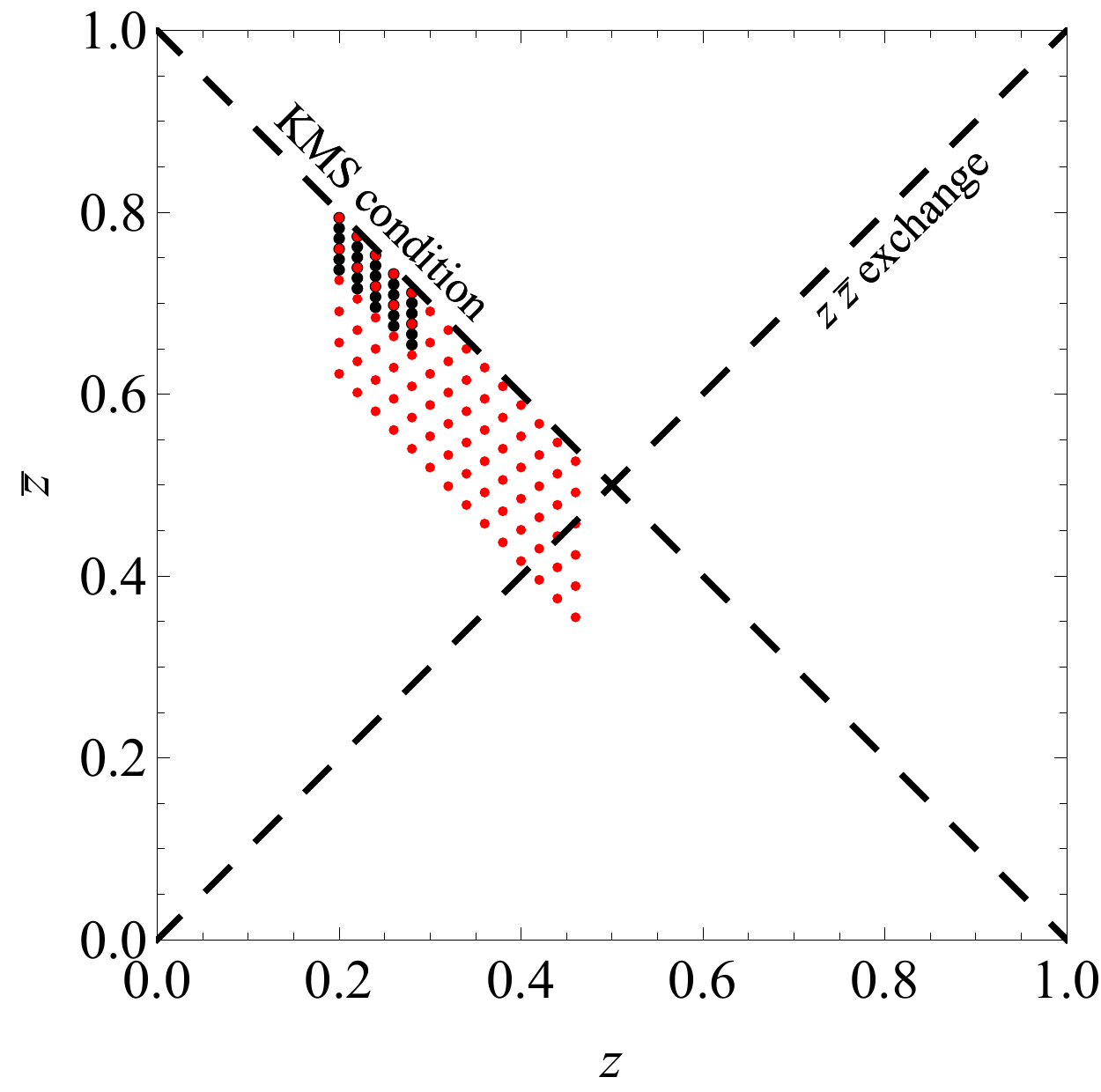}}
\caption{Example of the smallest (black) and largest (red) regions in ($z$, $\bar z$) where we minimize the square of the difference of the two-point function and it's periodic image, as in~\eqref{eq:KMS-relation}.  }
\label{fig:zzbar region for KMS}
\end{figure}

Finally, we will input the thermal coefficients we've obtained for the three families $[\s\s]_0$, $[\s\s]_1$, and $[\e\e]_0$ into the $\langle\s\s\rangle$ correlator, and impose the KMS condition to determine the last unknown $a_\e^{\langle\s\s\rangle}$. We do this as follows. We evaluate the correlator minus its image  under crossing in various  regions of the $(z,\zb)$ plane, $\mathcal P_{\text{KMS}}$. To determine $a_\e^{\langle\s\s\rangle}$, we attempt to minimize:
\begin{align}
\label{eq:KMS-relation}
\Lambda_{KMS}(a_\epsilon^{\<\s\s\>}) = \sum_{(z, \bar z) \in \mathcal{P}_{\text{KMS}}}(g(z,\zb)-g(1-z,1-\zb))^2.
\end{align} 
By setting $\partial \Lambda_{KMS}(a_\epsilon^{\<\s\s\>})/\partial a_\epsilon^{\<\s\s\>} = 0$ we can finally determine the results obtained in \eqref{eq:final-results}.

 The thermal inversion formula guarantees that the KMS condition is satisfied in the proximity of the point $ (z, \bar z) = (0, 1)$. Thus, if one tries to approximately impose KMS solely in that region, there would be an almost flat direction associated to the unknown $a_\e^{\langle\s\s\rangle}$ and, consequently, our numerical estimates would be inaccurate. However, if one imposes KMS in a region where the OPE does not converge well the results would once again be inaccurate. Thus, we try to impose that KMS is approximately satisfied in an intermediate region and check for robustness under changes of $\mathcal P_{\text{KMS}}$ within this  intermediate regime. We find that our results are indeed  robust for various choices of the $(z, \bar z)$ region $\mathcal P_{\text{KMS}}$ and, as mentioned before, for the choice of $\bar z$ values $\cP_{\text{mix}}$ which are used to perform the mixing of the three families. To emphasize this, in figure \ref{fig:thermal coefficients of families}, we show a spread of the thermal coefficients obtained by minimizing \eqref{eq:KMS-relation} for the values of $\cP_{\text{mix}}$ in \eqref{eq:mixing-zbar-vals} and for values of $\mathcal P_{\text{KMS}}$ raging between the two regions showed in \eqref{fig:zzbar region for KMS}. While the value of $a_\epsilon^{\<\s\s\>}$ varies by at most $\sim 10\%$ between any two choices of $\cP_{\text{mix}}$ and $\mathcal P_{\text{KMS}}$, the thermal coefficients for all other operators exhibit a much lower variance.\footnote{This is partly due to the fact that the contribution of the unit operator dominates the thermal coefficients of higher spin operators.} For instance, the stress energy tensor thermal coefficient varies by $\sim 5\%$, while the the thermal coefficient of the spin-$4$ operator $[\s\s]_{0, \ell = 4}$ varies by $\sim 1\%$. To test how well the crossing equation is satisfied on the Euclidean thermal cylinder we plot the difference 
 \be \label{eq:delta-KMS} \delta g_{\text{KMS}}(\tau, x) = g(x, 1+\tau) - g(x, \tau)\,,\ee
  in figure \ref{fig:diff-crossing}. The KMS condition is very close to being satisfied in the regime in which both the points $(x, \tau)$ and $(x, 1+\tau)$ are close to the origin of the s-channel OPE, $(0, 0)$.  For instance, we find that $\delta g_{\text{KMS}}(-1/4, 1/4)/g(-1/4, 1/4)= 0.0037$. This shows the great extent through which one could use the thermal inversion formula to systematically solve the KMS condition or, equivalently, solve the ``crossing-equation" of  El-Showk and Papadodimas  \cite{ElShowk:2011ag}.

\section*{Acknowledgements}

We thank Raghu Mahajan and Eric Perlmutter for collaboration in the early stages of this project and many stimulating discussions on finite-temperature physics. We also thank M.~Hasenbusch for providing useful references and for sharing unpublished Monte-Carlo results through private correspondence. We additionally thank Tom Hartman and Douglas Stanford for discussions. DSD and MK are supported by Simons Foundation grant 488657 (Simons Collaboration on the Nonperturbative Bootstrap), a Sloan Research Fellowship, and a DOE Early Career Award under grant No.~DE-SC0019085. LVI is supported by Simons Foundation grant 488653.

\appendix

\section{Details of the Monte-Carlo simulation}
\label{app:mc}

To compute the thermal two-point function $\<\s\s\>_\b$ using Monte-Carlo integration, we implemented Wolff's cluster algorithm on a periodic square lattice of size $40\x500\x500$. We used the spin-spin coupling $\b_\mathrm{critical} = 0.22165463(8)$ from \cite{Hasenbusch:2011yya}. The periodic direction of size $40$ represents the thermal circle, while the directions of size 500 approximate noncompact $\R^2$. The MC integration was performed over $4 \times 10^8$ iteration steps.  

As usual, there are three main sources of error: statistical error, finite-size effects (IR), and lattice-size effects (UV).  One of the nice properties of thermal correlators is that finite-size effects are much easier to control than for flat-space correlators. The reason is that we can imagine dimensionally reducing our system along the thermal circle. The result is a theory with thermal mass $m_\mathrm{th}\sim 1/\b$, and consequently fluctuations in the noncompact directions die off like $e^{-x/\b}$. Thus, on a torus with lengths $\b\x L \x L$, we expect corrections from the finiteness of $L$ to be suppressed by $e^{-L/\b}\sim 4\x10^{-6}$. By contrast, to compute flat-space two-point functions, one must consider torii with size $L\x L \x L$. In that case, finite-size effects go like $(L/x)^{-\De_\cO}$, where $\cO$ is the leading operator appearing in the OPE.

Thus, we expect that finite-size effects are negligible. Our main sources of error are statistical (visible as jitteriness in figure~\ref{fig:2pt function plots}) and lattice effects which cause the simulation to become inaccurate near the coincident point singularity.

\section{Sums over families of operators - $\alpha$ sums} 
\label{appendix: alpha sums}

Let's recall how to evaluate sums over a family of operators in the OPE of the thermal two-point function. The $t$-channel sum over a family $f$ consists of sums like
\begin{align}	
\sum_{\ell} \frac{d \bar h}{d\ell} S_{c,\Delta}(\bar h) (1-z)^{h_f+\delta(\bar h)-h_e} (1-\bar z)^{\bar h-h_e}\, ,
\end{align} where $\bar h = h_f + \ell + \delta(\bar h)$, $h_f$ is half the twist of a family $f$, and $h_e$ is the total $h$ of the external operators. Expanding in small $\delta(\hb) \log (1- z)$,
\begin{align}
\sum_{\ell} \frac{d \bar h}{d\ell} S_{c,\Delta} (\bar h) (1-\bar z)^{\bar h-h_e} \sum_{m=0}^\infty \frac{\delta(\bar h)^m}{m!} \log^m (1-z) (1-z)^{h_f-h_e},\label{eq:double-twist-sum-delta-expansion}
\end{align}
the sums we need to evaluate are of the form
\begin{align}
\sum_{\ell} \frac{d \bar h}{d\ell} p(\bar h) (1-\bar z)^{\bar h - h_e} %= \sum_{a\in A} c_a \bar z^a + \sum_{k=0}^{\infty} \alpha_k \bar z^k,
\label{eq:sum to compute}
\end{align} for a class of functions $p(\hb)$. The sum should be of the form
\begin{align}
\sum_{\ell} \frac{d \bar h}{d\ell} p(\bar h) (1-\bar z)^{\bar h - h_e} = \sum_{a\in A} c_a \bar z^a + \sum_{k=0}^{\infty} \alpha_k \bar z^k,
\label{eq:reg and sing part of sum}
\end{align} with $A\subset \mathbb{R}\backslash \mathbb{Z}_{\ge0}$.
The task is to compute the coefficient $c_a$ and $\alpha_k$. First, using the analytic expressions for $\delta(\hb)$, we determine the large-$\bar h$ asymptotics of $p(\bar h)$ in terms of the known functions $S_{a,\Delta}(\bar h)$,
\begin{align}
p(\bar h) \sim \sum_{a\in A} c_{a,\Delta}[p] S_{a,\Delta}(\bar h). \label{eq:S asymptotics of summand}
\end{align} 
The main idea is to use the integer-spaced sum\footnote{In this section we will write $S_{c,\Delta}(\hb)$ to denote the function with $z_\text{max}=\infty$. The difference with the finite $z_\text{max}$ is exponentially decaying at large $\hb$, and therefore does not contribute to the asymptotics and can be treated separately from the $z_\text{max}=\infty$ piece.} 
\begin{align}
\sum_{\substack{\bar h={\bar h_0} + \ell \\ \ell=0,1,\dots}} S_{a,\Delta}(\bar h) (1-\bar z)^{\bar h} &= (1-\bar z)^{{\bar h_0}} S_{a,\Delta}({\bar h_0}) \PFQ{2}{1}{1,{\bar h_0}-\Delta-a}{{\bar h_0}-\Delta+1}{1-\bar z}
\nn\\ &= \bar z^a (1-\bar z)^\Delta - S_{a-1,\Delta+1} ({\bar h_0}) (1-\bar z)^{{\bar h_0}} \PFQ{2}{1}{1,{\bar h_0}-\Delta-a}{-a+1}{\bar z}\label{eq:integer-spaced-sum}
\end{align} to determine the coefficients $c_a$ in \eqref{eq:reg and sing part of sum} in terms of the asymptotics $c_{a,\De}[p]$. 
Then, to compute the remaining terms that are regular in $\zb$, we regulate the sum \eqref{eq:reg and sing part of sum} by subtracting the sum in \eqref{eq:integer-spaced-sum} for each asymptotic of $p(\hb)$ in \eqref{eq:S asymptotics of summand}. With the asymptotics controlled, expanding the summand in small $\zb$ gives convergent sums in $\hb$ for the $\alpha_k$ coefficients. 
The $\alpha$ sum can be evaluated by the formula
\begin{align}
\alpha_k [p,\delta,h_e]({\bar h_0}) &= -\oint_{h_c -i\infty}^{h_c + i\infty} \frac{d\bar h}{2\pi i}  \genfrac{(}{)}{0pt}{}{\bar h - h_e}{k} (-1)^k \nonumber
\\ & \qquad \times \left(\pi \cot(\pi(\bar h-{\bar h_0}-\delta(\bar h))) ~ p(\bar h) - \pi \cot(\pi(\bar h -{\bar h_0})) \sum_{\substack{a\in A \\ a<K}} c_{a,\Delta} S_{a,\Delta}(\bar h) \right) \nonumber
\\ & \quad + \sum_{\substack{a\in A \\ a<K}} c_{a,\Delta} \left(r_k(a,\Delta,h_e,{\bar h_0}) +s_k(a,\Delta,h_e,{\bar h_0})\right). \label{eq:def:alpha_k}
\end{align}
Here, $K$ should be at least $k$, but larger $K$ gives a faster converging integral. The contour is at $h_c = \hb_0+\delta(\hb_0)-\epsilon$. In the last line, we have added back terms with $r_k$, which is the coefficient of $\bar z^k$ for the integer spaced sum in \eqref{eq:integer-spaced-sum},
\begin{align}
r_k(a,\Delta,h_e,{\bar h_0}) &= \left.- S_{a-1,\Delta+1} ({\bar h_0}) (1-\bar z)^{{\bar h_0}-h_e} \PFQ{2}{1}{1,{\bar h_0}-\Delta-a}{-a+1}{\bar z} \right|_{\bar z^k}
\nn\\ &= - S_{a-1,\Delta+1} ({\bar h_0}) \sum_{m=0}^k (-1)^m \genfrac{(}{)}{0pt}{}{{\bar h_0} - h_e}{m} \frac{({\bar h_0}-\Delta-a)_{k-m}}{(-a+1)_{k-m}} \label{eq:def:regular-terms}
\end{align} and $s_k$, which is the contribution of spurious poles (coming from the asymptotics $S_{a,\Delta}(\bar h)$ we subtracted) that are picked up by the contour when $h_c -\Delta - a \le 0$,
\begin{align}
&s_k(a,\Delta,h_e,{\bar h_0}) = \sum_{n=0}^{\lfloor a+\Delta-h_c\rfloor} \operatorname*{Res}\limits_{\bar h= a+\Delta-n} \genfrac{(}{)}{0pt}{}{\bar h - h_e}{k} (-1)^k \pi \cot (\pi(\bar h -{\bar h_0}) )S_{a,\Delta}(\bar h)
\nn\\
&\qquad = \sum_{n=0}^{\lfloor a+\Delta-h_c\rfloor} \genfrac{(}{)}{0pt}{}{a+\Delta-n - h_e}{k} (-1)^k \pi \cot (\pi(a+\Delta-n -{\bar h_0})) \frac{(-1)^n}{n! \Gamma(-a) \Gamma(a-n+1)}.\label{eq:def:spurious-poles}
\end{align} The contour integral can be integrated numerically to high precision.

\bibliography{Biblio}{}
\bibliographystyle{utphys}

\end{document}